\documentclass{article}
\usepackage{a4wide} 
\usepackage{bm} 
\usepackage{amssymb}
\usepackage{natbib}
\usepackage{graphicx}
\newcommand{\ee}{\end{equation}}
\newcommand{\be}{\begin{equation}}
\newcommand{\la}{\langle}
\newcommand{\ra}{\rangle}
\newcommand{\dt}{\delta_{t}}
\newcommand{\dx}{\delta_{x}}

\newcommand{\hi}{\bar{l}}
\newcommand{\hj}{\bar{k}}

\newcommand{\br}{{\bf r}}    
\newcommand{\bv}{{\bf v}}    
\newcommand{\bu}{{\bf u}}    
\newcommand{\bc}{{\bf c}}

\begin{document}
\bibliographystyle{jfm}
\title{Mesoscopic modeling of heterogeneous boundary conditions for
  microchannel flows}

\author{R. Benzi$^1$, L. Biferale$^{1}$, M.  Sbragaglia$^1$, S.
  Succi$^{2}$ and F. Toschi$^{2}$\\
$^1$ Dipartimento di Fisica, Universit\`a ``Tor
  Vergata'', and INFN,\\Via della Ricerca Scientifica 1, I-00133 Roma,
  Italy.\\ $^2$ CNR, IAC, Viale del Policlinico 137, I-00161 Roma,
  Italy.}

\maketitle

\begin{abstract}
  We present a mesoscopic model of the fluid-wall interactions for
  flows in microchannel geometries. We define a suitable
  implementation of the boundary conditions for a discrete version of
  the Boltzmann equations describing a wall-bounded single phase
  fluid.  We distinguish different slippage properties on the surface
  by introducing a slip function, defining the local degree of slip
  for mesoscopic {\it molecules} at the boundaries.  The slip function
  plays the role of a {\it renormalizing} factor which incorporates,
  with some degree of arbitrariness, the microscopic effects on the
  mesoscopic description.  We discuss the mesoscopic slip properties
  in terms of slip length, slip velocity, pressure drop reduction
  (drag reduction), and mass flow rate in microchannels as a function
  of the degree of slippage and of its spatial distribution and
  localization, the latter parameter mimicking the degree of roughness
  of the ultra-hydrophobic material in real experiments. We also
  discuss the increment of the slip length in the {\it transition
  regime}, i.e.  at ${\cal O}(1)$ Knudsen numbers.  Finally, we
  compare our results with Molecular Dynamics investigations of the
  dependency of the slip length on the mean channel pressure and local
  slip properties (Cottin-Bizonne {\em et al.} 2004) and with
  the experimental dependency of the pressure drop reduction on the
  percentage of hydrophobic material deposited on the surface -- Ou
  {\em et
  al.} (2004).
\end{abstract}

\section{Introduction}
The physics of molecular interactions at fluid-solid interfaces is a
very active research area with a significant impact on many emerging
applications in material science, chemistry, micro/nano\-engineering,
biology and medicine, see \cite{white}, \cite{elettro}, \cite{micro},
\cite{MEMS}.  As for most problems connected with surface effects,
fluid-solid interactions become particularly important for micro- and
nano-devices, whose physical behaviour is largely affected by high
surface/volume ratios.  Recently, due to an ever-increasing interest
in microfluidics and MEMS (micro-electromechanical system)-based
devices, experimental capabilities to test and analyze such systems
have undergone remarkable progress.

In this paper, we shall focus on flows in micro-channels, a subject
which has recently become accessible to systematic experimental
studies thanks to the developments of silicon technology (see
\cite{tabebook}, \cite{karniakadisbook} and references therein).

Classical hydrodynamics postulates that a fluid flowing over a solid
wall sticks to the boundaries, i.e. the fluid molecules share the same
velocity of the surface, \cite{batch,Gold}.  This law, and its
consequences, are well verified at a macroscopic level, where the
characteristic scales of the flow are much larger than the molecular
sizes.  The situation changes drastically at a microscopic level.
Many experiments (\cite{tabe,ou,wata,vino,vinorev,pit,baudry,craig,zhu1,zhu2,cheng,tretheway,bonna1,bonna2,choi,janus})
and numerical simulations using Molecular Dynamics
(\cite{MD1,MD2,MD3,MD4,MD5,MD6,MD7,MD8}) have shown evidences that the
solid-fluid interactions are strongly affected by the chemico-physical
properties and by the roughness of the surface.  For example, water
flowing over a hydrophilic, hydrophobic or super-hydrophobic surface,
may develop quite different flow profiles in micro-structures.  One of
the most spectacular effect is the appearance of an effective slip
velocity, $V_s$, at the boundary, which, in turn, may imply a
reduction of the kinetic energy dissipation with significant
enhancement of the overall throughput (at a given pressure drop), see
\cite{wata,vino,pit,baudry,craig,zhu1,zhu2,cheng,tretheway,bonna1,choi}.
From the slip velocity, one defines a slip length, $L_s$, as the
distance from the wall where the linearly extrapolated velocity
profile vanishes. The experimental and theoretical picture is still
under active development. No clear systematic trend of the slip effect
as a function of the chemico-physical components has been found to
date. Slip lengths varying from hundreds of $n$m up to tens of $\mu$m
have been reported in the literature.  Moreover, controversial claims
about the importance of the roughness of the surface and of the
combined degree of roughness-hydrophobicity have been presented. In
simple flows roughness is expected to increase the energy exchange
with the boundaries, inducing a corresponding decrease in the
slippage.  However, both increase and decrease of the slip length as a
function of the surface roughness have been claimed in the literature,
\cite{zhu1,zhu2,bonna1,bonna2}.
From a purely molecular point of view, a critical parameter governing
the solid-liquid interface is the contact angle (wetting angle).
Clean glass is highly hydrophilic, with an angle with water close to
$\theta= 0^o$ (perfect wetting).  Recently ultra-hydrophobic surfaces
have been obtained which prove capable of sustaining a contact angle
with water as high as $\theta=177^o$ a value at which water droplets
are almost spherical on the surface, \cite{chen,fadeev}.

Some authors proposed that the increase in the slippage might be due
to a rarefaction of the flow close to the wall, a depleted water
region or vapor layer should exists near a hydrophobic surface in
contact with water, \cite{Lum,Sak,Sch,tyrrel}.  Recent Molecular
Dynamics simulations have also presented some evidences of a {\it
dewetting} transition, leading to a strong increase of the slip
length, below some capillarity pressure in microchannels with
heterogeneous surface, \cite{barrat1,barrat}.  The physics looks very
fragile, depending as it seems on many complicated chemical and
geometrical details.

From the numerical point of view, Molecular Dynamics (MD)
 is the standard tool to systematically
investigate the problem, \cite{review1,review2,YIP}. In MD the solid-liquid and the liquid-liquid
interactions are introduced by using Lennard-Jones type potential
(with interaction energies and molecular diameters adjusted from
experiments).  By changing the interaction energies one may tune the
surface tension and, consequently, the contact angles.  MD also offers
the possibility to model the boundary geometries and roughness with a
high degree of fidelity.  The main limitation, however, is the modest
range of space and especially time-scales, which can be simulated at a
reasonable computational time, typically a few nanoseconds, \cite{review2,KOPLIK}.

The coupling between MD and hydrodynamic modes involves a huge gap of
space and time scales.  Recently, an interesting attempt to reproduce
MD simulations of heterogeneous microchannels with a continuum
mechanical description based on Navier-Stokes equations and suitable
hydrodynamic boundary conditions has been proposed,
(\cite{barrat1,barrat,MD8}).  \cite{barrat}  show that
some of the results obtained by MD simulations of a microchannel with
a grooved surface can be qualitatively reproduced using a Stokes
equation for the incompressible flow, in combination with an
heterogeneous boundary condition, linking the slip velocity parallel
to the boundary $\bu_{||}(\br)$ to the stress in the normal direction,
$\hat{n}$:

\begin{equation}
 \bu_{||}(\br) = b(\br) \partial_n \bu_{||}(\br)
\label{eq:bcslip}
\end{equation} 

where $b(\br)$ is a position dependent normalized slip length
mimicking the heterogeneity of the microscopic level. The qualitative agreement with the results of MD
simulations can be obtained by properly tuning the $b(\br)$ values.
In particular, they show that the dewetting transition observed in MD
simulations, for some values in the Pressure-Volume diagram, is
equivalent to the assumption, at the hydrodynamic level, that the
boundary surface is made up of alternating strips of free-shear (high
slip length $b(\br)$) and {\it wetting} material (low slip length).

In this paper, we main aim at filling the gap between the microscopic
description typical of MD, and the macroscopic level of the
Navier-Stokes equations by using a mesoscopic model based on the
Boltzmann Equation. In particular, we will use a discrete model known
as Lattice Boltzmann Equation (LBE) with heterogeneous boundary
conditions.

The boundary condition (\ref{eq:bcslip}), \cite{Max}, arises naturally
in a power expansion of the Boltzmann equation in terms of the Knudsen
number, $Kn=\lambda/L$, that is, the ratio between the mean free path,
$\lambda$, and a typical length of the channel, $L$.  At first order
in $Kn$, one obtains the Navier-Stokes equation with the Maxwell
boundary conditions above, \cite{cerci,HADJ}. Still, recent
experimental results raised some doubts on the validity of this
construction above some critical value of the Knudsen number,
\cite{tabe}. There, the authors report that above $Kn \sim 0.3 \pm
0.1$ both helium and nitrogen exhibit a non linear dependence of the
flow rate on $Kn$ which cannot be explained by solving the Stokes
equation with the first order slip boundary condition
(\ref{eq:bcslip}). For those values of $Kn$, the flow is in the
so-called {\it transition regime} and it has been shown that the
coupling between hydrodynamic equations with a second order boundary
condition
\begin{equation}
 \bu_{||}(\br) = b_1(\br) \partial_n \bu_{||}(\br) + b_2(\br) \partial^2_n \bu_{||}(\br)
\label{eq:bcslip2}
\end{equation}
is more appropriate to fit the experimental data \cite{tabe}.

The purpose of our investigation is twofold.  First, we aim at
developing a model which allows a coarse-grained treatment of local
effects close to the flow-surface region, without delving into the
detailed molecule-molecule description typical of MD.  Second, we wish
to design a tool capable of describing fluid motion also beyond the
{\it linear Knudsen regime}.

The underlying hope behind the present hydro-kinetic approach, is
that the main features of the fluid-surface interactions can be
rearranged into a suitable set of {\it renormalized} LBE boundary
conditions. This implies that all details of the contact angle, the
solid-fluid interaction length, the local microscopic degree of
roughness, can, to some extent, be included within the local
definition of effective accommodation factors governing the
statistical interactions between the mesoscopic {\it molecules} and the
solid walls, \cite{PRLSLIP,ss04}.  In a more microscopic vein, one may
also describe the interactions between solid-liquid and liquid-liquid
populations using a {\it mean-field} multi-phase LBE description, see
\cite{BE2phase1,BE2phase2,BE2phase3,BE2phase4,BE2phase5,KWOK}.
Results based on these more sophisticated schemes will be reported in
a forthcoming paper, \cite{bbsst05}.

The paper is organized as follows. In section (\ref{sec:LBE}) we
briefly remind the main ideas behind the lattice versions of the
Boltzmann equations and we present a natural way to implement
non-homogeneous slip and no-slip boundary conditions in the model.  In
section (\ref{sec:hydro}) we discuss the hydrodynamic limit of the LBE
previously introduced with particular emphasis on the form of the
hydrodynamic boundary conditions in presence of slippage.  In section
(\ref{sec:num}) we present the numerical results at various Knudsen
and Reynolds numbers, as well as a function of the degree of slippage
and localization.  Whenever directly applicable we compare the results
obtained within our mesoscopic approach with (i) exact results in the
limit of small Knudsen numbers obtained in the hydrodynamic formalism,
\cite{Phi1,Phi2,stone} (ii) results obtained with a microscopic
approach using MD simulations, \cite{barrat} and (iii) recent experimental results
of microchannels with ultrahydrophobic surfaces, \cite{ou}.  Conclusions
and perspectives follow in section (\ref{sec:conc}).  Technical
details are given in the appendices.

\section{Lattice Kinetic formulation}
\label{sec:LBE}
The Boltzmann Equation describes the space-time evolution of the
probability density $f(\br,\bv,t)$ of finding a particle at position
$\br$ with velocity $\bv$ at a given time $t$. This evolution is
governed by the competition between free-particle motion and molecular
collisions which promote relaxation towards a non-homogeneous
equilibrium, whose distribution $f^{eq}(\rho,\bu)$, is the Maxwellian
consistent with the local density, $\rho(\br)$, and coarse grained
velocity, $\bu(\br)$. The hydrodynamic variables are obtained as
low-order moments of the velocity distributions.  Infact, the
hydrodynamic density and velocity are $\rho(\br,t) = \int d\bv
f(\br,\bv)$, and $\bu(\br,t) = \int d\bv \bv f(\br,\bv)$,
respectively. The Navier-Stokes equations for the hydrodynamic fields
are recovered in the limit of small-Knudsen numbers using the
Chapman-Enskog expansion \cite{cerci2}.  The Boltzmann equation lives
in a six-dimensional phase-space and consequently its numerical
solution is extremely demanding, and typically handled by stochastic
methods, primarily Direct Simulation Monte Carlo (for a review see \cite{BIRD}).
However, in the last fifteen years, a very appealing alternative (for
hydrodynamic purposes) has emerged in the form of lattice versions of
the Boltzmann equations in which the velocity phase space is
discretized in a minimal form, through a handful of properly chosen
discrete speeds (of order ten in two dimensions and twenty in three
dimensions --see appendix A for details).

\begin{figure*}
\begin{center}
\includegraphics[scale=0.4]{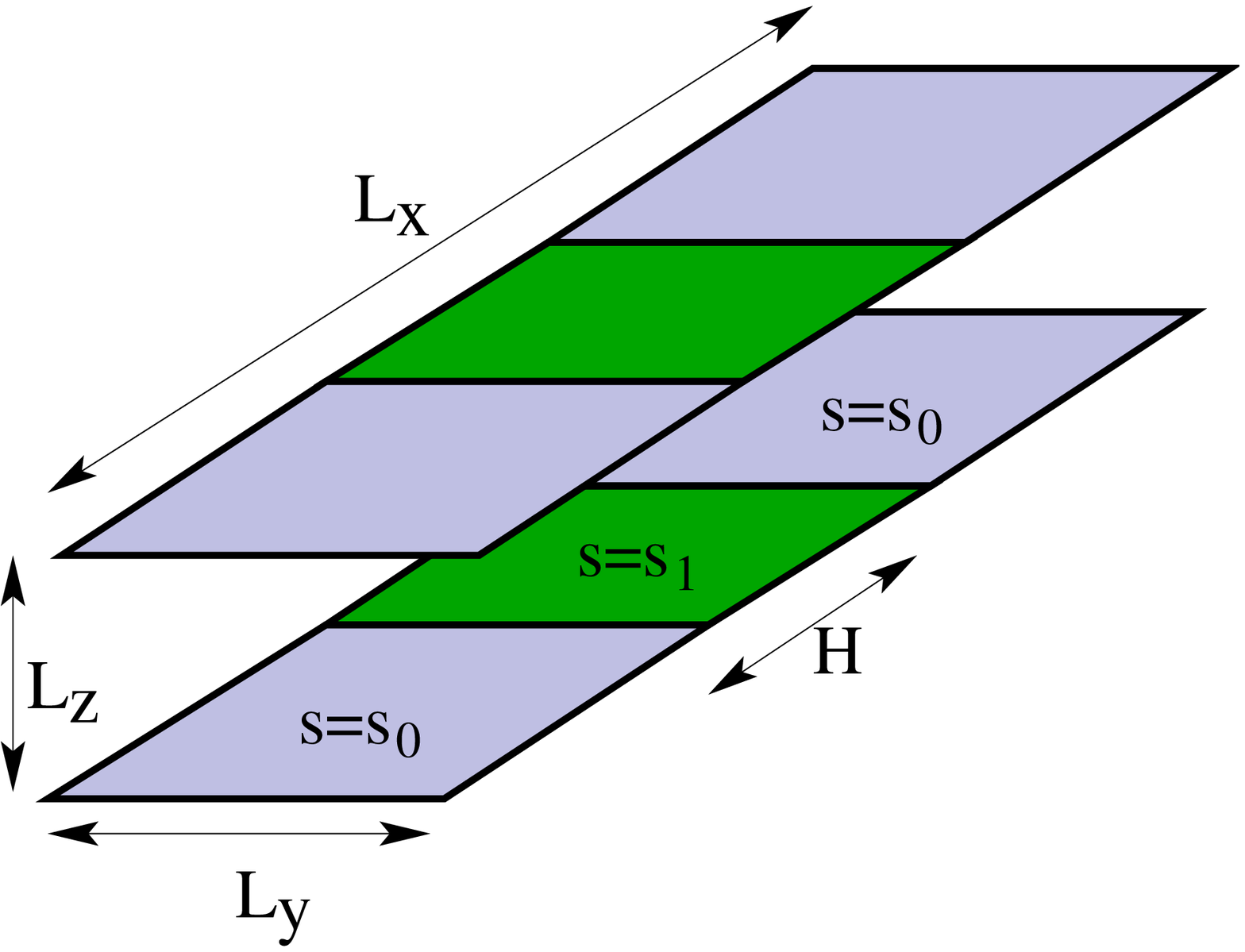}
\includegraphics[scale=0.4]{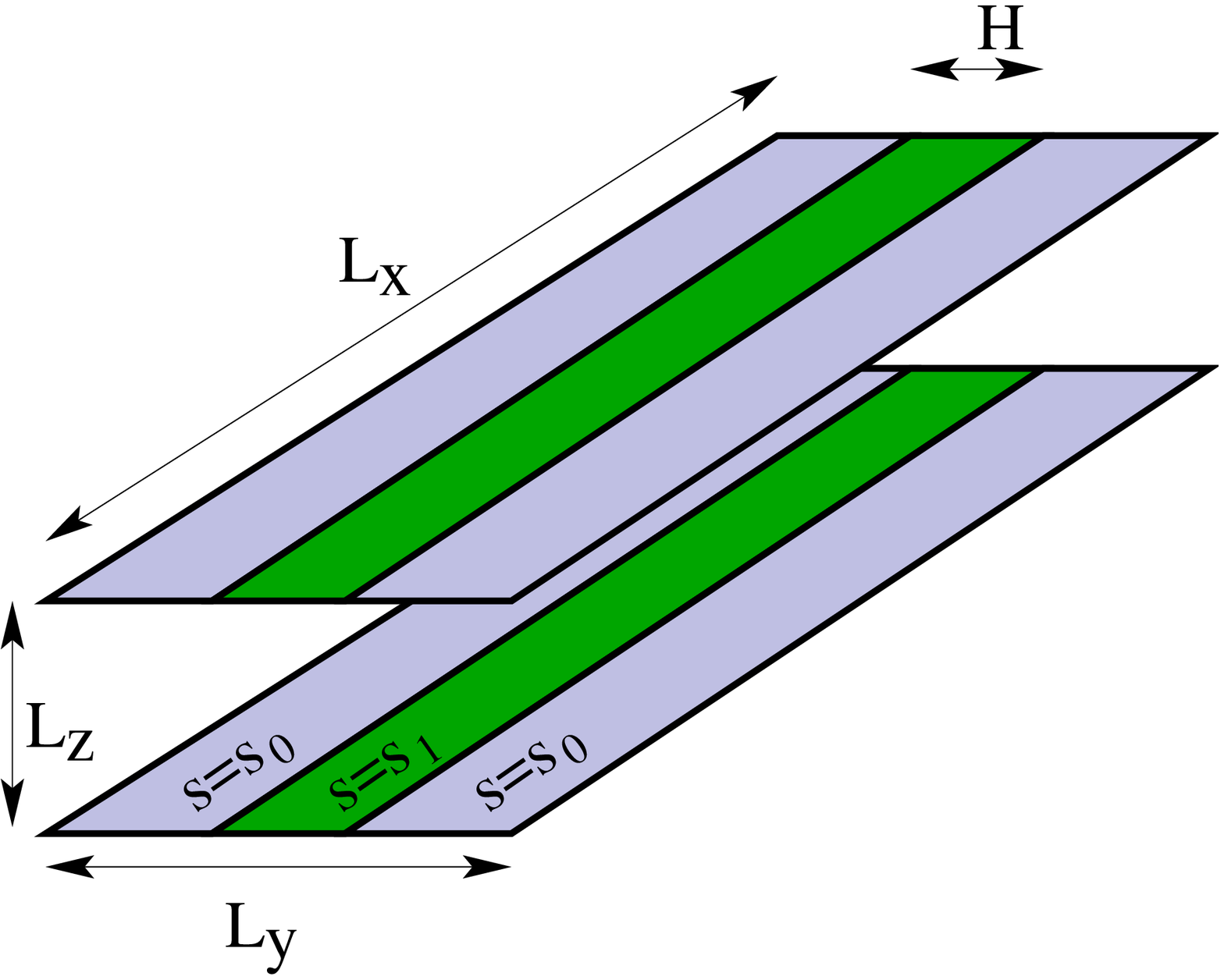}
\caption{Typical geometry of the microchannel configuration.  We have
  periodic boundary conditions along the stream-wise, $\hat{x}$, and
  span-wise $\hat{y}$ directions. The two rigid walls at $z=0,L_z$ are
  covered by two strips of width $H$ and $L-H$, where $L=L_x$ for
  transversal strips (left panel) and $L=L_y$ for longitudinal strips
  (right panel).  The two strips have different slippage properties
  identified by the values $s_0$ and $s_1$.  The ratio $\xi=H/L$
  identifies the fraction of hydrophobic material deposited on the
  surface. Typical sizes used in the LBE simulations are $L_x=L_y=64$
  grid points and $L_z=84$ grid points. This would correspond, for
  example, for a fluid like water at $Kn=10^{-3}$, to a microchannel
  of height of the order of $100 \,\mu$m.}
\label{fig:configuration} 
\end{center}
\end{figure*}

This leads to the Lattice Boltzmann Equations (LBE)
 for the probability
density, $f_l(\br,t)$, where $\br$ runs over the discrete lattice, and
the subscript $l=0,N-1$ labels the $N$ discrete velocities values
allowed by the scheme, $\bv \in \{\bc_0,\cdots \bc_{N-1} \}$, \cite{Saurobook,gladrow,BSV,chendol,zanetti}. It is
interesting to remark that it is sufficient to retain a limited
numbers of discretized velocities at each site to recover the
Navier-Stokes equations in the hydrodynamic limit.  In two dimensions
the nine-speed $2DQ9$ model ($N=9$) is in fact one of the most used
2d-LBE scheme, due to its enhanced stability \cite{KARLI}.  All
three-dimensional simulations described in this paper are based on the
the $3DQ19$ scheme ($N=19$) (see Fig. \ref{fig00} in appendix A for a
graphical description of LBE velocities in 2d and 3d).  For the sake
of concreteness, we shall refer to the two-dimensional nine-speed
$2DQ9$ model, although the proposed analysis can be extended in full
generality to any other discrete-speed model living on a regular
lattice.  We begin by considering the Lattice Boltzmann Equation in
the following BGK approximation, \cite{BGK}:
\begin{equation}
\label{Boltzmann}
f_{l}(\br+\bc_{l},t+1)-f_{l}(\br,t)=
-\frac{1}{\tau}\left(f_{l}(\br,t)-f^{(eq)}_{l}(\rho,\bu)\right)+F_{l}
\end{equation}
where we have assumed lattice units $\delta_x=\delta_t =1$. In
(\ref{Boltzmann}), $\tau$ is the relaxation time to the local
equilibrium, which is proportional to the Knudsen number.  The
explicit expression of the speed vectors, $\bc_l$, of the lattice
equilibrium distribution, $f^{(eq)}_{l}(\rho,\bu)$ and of the forcing
term $F_l$ needed to reproduce a constant pressure drop, are described
in the appendix A.  The hydrodynamic fields in the lattice version are
expressed by:
\begin{equation}
\rho(\br)=\sum_{l}  f_{l}(\br);\,\,\, 
\rho(\br) \bu(\br)=\sum_{l} \bc_{l} f_{l}(\br).
\label{hlbe}
\end{equation} 
Boundary conditions for Lattice Boltzmann simulations of microscopic
flows have made the object of much investigation in recent years,
\cite{toschi,Niu,Lim,Ansumali}. In particular, we are interested in
studying the evolution of the LBE in a microchannel with heterogenous
boundary conditions (H-LBE) --the simplest case being a sequence of
two alternating strip with different slip properties, as depicted in
Fig. (\ref{fig:configuration}).  A general way of imposing the
boundary conditions in the LBE reads as follows: \be
f_{\hj}(\br_w,t+1)=\sum_{\hi} B_{\hj,\hi}(\br_w)f_{\hi}(\br_w,t) \ee
where the matrix $B_{\hj,\hi}$ is the discrete analogue of the
boundary scattering kernel expressing the fluid-wall interactions.
Here and in the following, we use the notation $\br_w$ to indicate the
generic spatial coordinate over the surface of the wall and the
indices $\hi,\hj$ label the subset of incoming and outgoing velocities
respectively.  To guarantee conservation of mass and normal momentum,
the following sum-rule applies: \be
\label{norm} \sum_{\hj} B_{\hj,\hi}(\br_w) =1.
\ee 

Upon the assumption of fluid stationarity, we can drop the dependence
on $t$ and write:

\be
\label{boundary}
f_{\hj}=\sum_{\hi} B_{\hj,\hi} (\br_w)f_{\hi}.  
\ee 

The simplest, non trivial, application involves a {\it slip function},
$s(\br_w)$, representing the probability for a particle to slip
forward, (conversely, $1-s(\br_w)$ will correspond to the probability
for the particle to be bounced back). If we focus, for example, on the
north-wall boundary condition (see Fig. (\ref{fig00}) in appendix A),
the boundary kernel upon the assumption of preserved density and zero
normal component of the velocity field (\ref{norm}) takes the form :
\be
\label{matrix} 
 \left( 
\begin{array}{c} 
 f_7 \\ f_4 \\ f_8 \end{array} \right)
= \left( \begin{array}{c c c} 1-s(\br_w) & 0 & s(\br_w) \\ 0
& 1 & 0 \\ s(\br_w) & 0 & 1-s(\br_w) \end{array} \right) 
 \left( 
\begin{array}{c} 
 f_5 \\ f_2 \\ f_6 \end{array} \right).
\ee
In this language, the usual no-slip boundary conditions are recovered
in the limit $s(\br_w) \rightarrow 0$ everywhere (incoming velocities
are equal and opposite to the outgoing velocities), while the perfect
free-shear profile is obtained with $s(\br_w) \equiv 1$.  The
formalism is sufficiently flexible to allow the study of both spatial
inhomogeneity of a given hydrophobic material and/or the effects
of different degrees of hydrophobicity at different spatial locations.

The above LBE scheme has been already successfully tested in the case
of a homogeneous slippage $s(\br_w) = s_0,\; \forall \br_w$,
\cite{ss04}.  In that case, it has been shown (see appendix B) that
the LBE scheme converges to an hydrodynamic limit with the slip
boundary condition
\begin{equation}
 \bu_{||}(\br_w) =A \, Kn\,  |\partial_n \bu_{||}(\br_w)| + B\, Kn^2\, |\partial^2_n \bu_{||}(\br_w)|,
\label{slip2}
\end{equation}
where the parameters $A,B$ can be tuned by changing the degree of
slippage, $s_0$ and the external forcing. In this case, the LBE reproduces the analytical
prediction for the  slip length, obtained by assuming the
existence of a Poiseuille velocity profile and, with a suitable
choice of $A,B$ in (\ref{slip2}), one can show that the
model is also able to fit the experimental non-linear dependencies on
the Knudsen number observed in \cite{tabe} for nitrogen and helium.
\section{Hydrodynamic limit}
\label{sec:hydro}
To begin with, we wish to analyze the hydrodynamic limit, $Kn
\rightarrow 0$, of the previous LBE models with non-homogeneous
boundary conditions, as dictated by the space-dependent profile of the
slip function, $s(\br_w)$, at the walls.  For the sake of simplicity,
we shall confine our attention to the continuum limit of zero lattice
spacing and time increments, $\delta_x=\delta_t \rightarrow 0$,
$c=\frac{\delta_x}{\delta_t} \rightarrow 1$.  Starting from the
discretized equations (\ref{Boltzmann}) one gets for the continuum
limit of the LBE:

\be \partial_t f_l + (\bc_{l} \cdot {\bf \nabla})
f_{l}=-\frac{1}{\tau}\left(f_{l}-f^{(eq)}_{l}\right) + F_l.
\label{BC}
\ee 

In the following, we shall be interested in the case of stationary,
time-independent, solutions (small-Reynolds regime).  To this purpose,
we may formally write the solution of (\ref{BC}) by using the
time-independent Green's function:

\be
\label{bf1} 
f_{l}(\br)=\sum_{n=0}^{\infty} (-1)^{n} (\tau (\bc_{l} \nabla))^{n} \left[
f^{(eq)}_{l}(\rho,\bu) + \tau F_l \right].  
\ee 

Let us notice that defining $\tau = \frac{Kn L_{z}}{c_{s}}$ ($c_{s}$
being the sound speed velocity), the above expression can be
interpreted as a formal solution in powers of the Knudsen number.  By
recalling the expression of the hydrodynamic fields (\ref{hlbe}), it
is readily checked that the boundary velocity can be expressed as a
function of the velocity stress, $\partial_i u_j$, at the boundary
itself. For sake of simplicity, we report here only the first order
term (in the Knudsen and Mach numbers) of the expansion (see Appendix
B):

\be
\label{BBB} 
\bu_{||}(\br_w)= Kn \left ( \frac{c}{c_{s}} \right )
\frac{s(\br_w)}{1-s(\br_w)} \left|\partial_n \bu_{||}(\br_w)\right| + {\cal O}(Kn^2)  
\ee 

which is a direct generalization of the result obtained 
for the case of homogeneous boundary conditions (\ref{slip2}).  The
main difference is that, due to the spatial dependence of the stress
tensor along the wall, subtle non-linear effects may be triggered by
the spatial correlation between the slip function $s(\br_w)$ and the
stress at the wall.

The hydrodynamic equations of motion in the 
stationary case read as follows: 
\be \left \{ \begin{array} {l} (\bu
\cdot \nabla )\bu = - \frac{{\bf \nabla }P}{\rho} + 
\frac{1}{\rho} \nabla \cdot (\nu \rho \nabla \bu)\\ 
\nabla \cdot
(\rho \bu) = 0 \\ 
\bu_{||}(\br_w)= Kn \left ( \frac{c}{c_{s}} \right )
\frac{s(\br_w)}{1-s(\br_w)} |\partial_n \bu_{||}(\br_w)| + {\cal O}(Kn^2) 
\\ \bu_{\bot} (\br_w)=0 \end{array} \right.
\label{hydro}
\ee where ${\bf \nabla} P$ contains both the imposed mean pressure
drop, ${\bf F}$, and the fluid pressure fluctuations.  In the limit of
small Mach numbers ($\frac{\Delta \rho}{\rho} \ll 1 $) we may take a
constant density $\rho=1$.  Let us notice that in this limit, the
incompressibility constraint $\nabla \cdot \bu =0$ imposes that any
non-homogeneity of $\bu_{||}$ along the wall-parallel direction must
be compensated by an equal and opposite gradient of the normal
velocity $\bu_{\bot}$. This implies that the local velocity profile
cannot be of {\it Poiseuille type} everywhere ($\bu_{\bot} =0$).

\begin{figure*}
\begin{center}
\includegraphics[scale=0.55]{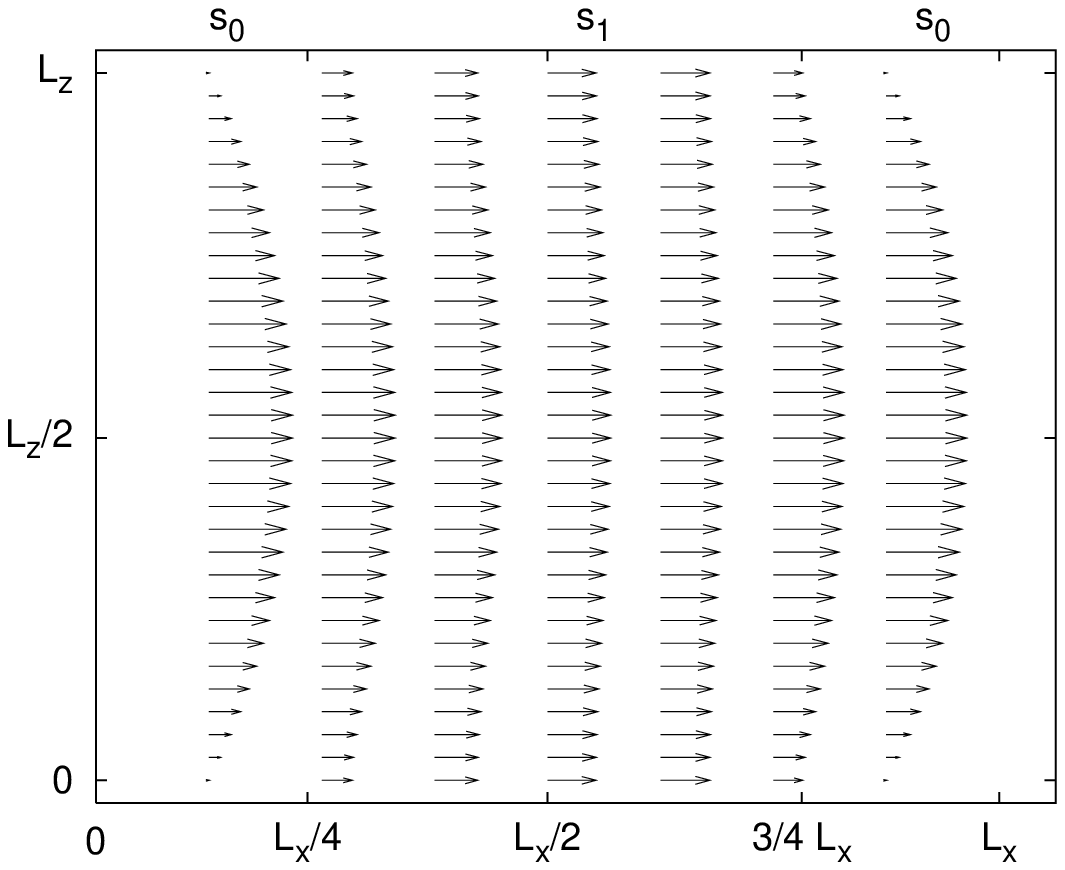}
\includegraphics[scale=0.55]{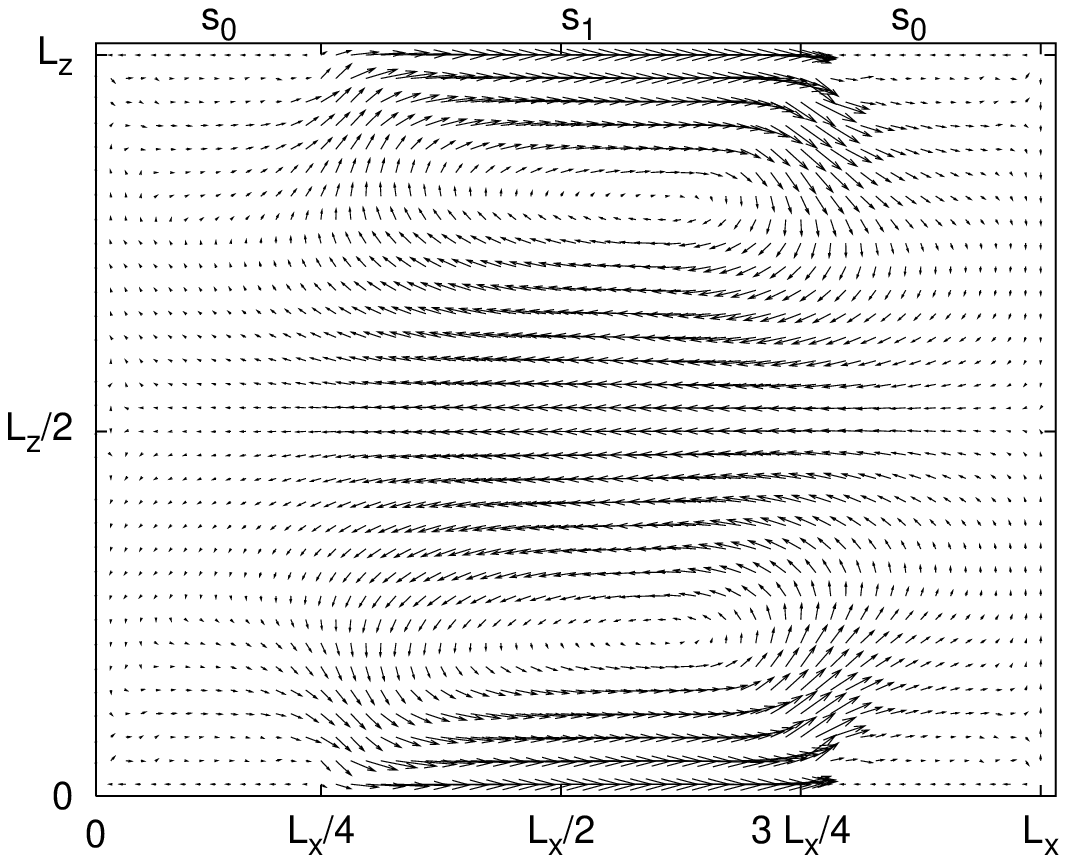}
\caption{Results in the plane $y=L_y/2$ along the channel measured in
  the transversal strip configuration (see left panel of Fig.
  \ref{fig:configuration}). The left panel shows the velocity profile.
  Notice that the pure inlet Poiseuille flow becomes an almost perfect
  shear free profile in the region with $s_1=1$. The right panel is
  meant to highlight the local differences between the pure Poiseuille
  flow and the measured profiles, showing the result for
  $(\bf{u_x(\br)}-\bf{u^{pois}_x(\br)})$. Notice the recirculation
  area, entering deep in the channel bulk, produced by the alternating
  slip and no-slip boundary conditions.}
\label{fig:prof}
\end{center}
\end{figure*}

In order to assess the effects of the slip on the global quantities,
it is useful to define the mean profile, $\la \bu (z) \ra$.  Let us
consider for instance the geometry depicted in Fig.
(\ref{fig:configuration}), where the direction perpendicular to the
walls is denoted by $\hat{z}$. We define an homogeneous mean profile
as:

\be
\la \bu (z) \ra = \frac{1}{S}\label{int} \int \bu (\br) \, dx dy 
\label{mhp}
\ee 

where $\la \cdots \ra$ stands for averaging over a plane parallel to
the boundary surface, $S$. Even though the local velocity does not
reproduce a Poiseuille profile, it can be shown from (\ref{hydro})
that in the case of periodic boundary conditions between inlet and
outlet flows, the mean homogeneous profile (\ref{mhp}) cannot develop
non linear stresses, namely: \be
\label{pois}  
\la \bu (z) \ra = u_{pois}(z) + u_{slip} 
\ee

with the notable fact that a slip velocity may appear at the boundary.
In the above definition, (\ref{pois}) $u_{pois}(z)$ stands for the
Poiseuille parabolic profile with zero velocity at the boundary. A
first set of qualitative results are plotted in Fig. (\ref{fig:prof}),
where the local velocity profiles and the difference between the
observed velocities and the standard no-slip Poiseuille flow are
shown.

From the expression (\ref{pois}), one may define a macroscopic, global
slip length, as the distance away from the wall at which the linearly
extrapolated slip profile (\ref{pois}) vanishes:

\be
L_s = \frac{u_{slip}}{|\partial_z u_{pois}(z_{w})|}
\label{slipL}
\ee

where $|\partial_z u_{pois}(z_{w})|$ is the Poiseuille stress
evaluated at the wall. Similarly, one may define the {\it mass flow
rate} gain $G$ as

\be 
G \equiv \frac{\Phi_{s}}{\Phi_{p}} = \left(1 + \frac{6L_s}{H}\right)
\label{eq:mfr}
\ee

being $\Phi_{s}=\int u_{x}(\br) dydz$ the real mass flow rate and
$\Phi_{p}$ the Poiseuille mass flow rate for our configuration:

\be
\Phi_{p}=\left( - \frac{d P}{d x} \right) \frac{ L^{3}_{z}L_{y}}{12 \mu }
\label{massflowgain}
\ee

with $\mu$ the dynamic viscosity of the fluid.  In terms of these
quantities one can define the pressure drop reduction,

\be
\Pi=\frac{\Delta P_{no-slip}-\Delta P}{\Delta P_{no-slip}},
\label{eq:pdr}
\ee

which is defined as the gain with respect to the pressure drop
corresponding to a non slip channel with the same overall throughput,
$\Phi_{s}$. The pressure drop reduction, $\Pi$, is usually
interpreted as an effective {\it drag reduction} induced by the slippage,
\cite{ou}.

\section{Numerical results}
\label{sec:num}
Next, we present the numerical results obtained from the H-LBE model
by changing the spatial distribution and intensity of the slip
function at the boundaries.  We shall also address dependencies of the
slip flow on the Knudsen and Reynolds numbers.

We begin by investigating the dependency of the macroscopic slip
length, $L_s$, and the average mass flow rate through the channel, on
the total amount of {\it slip material} deposited on the surface.  The
natural control parameter to investigate this issue is the average of
the slip function on the boundary wall:
\begin{equation}
s_{av} =  \la s(\br_w) \ra =\frac{1}{S} \int  s(\br_w) dS
\label{eq:mass}
\end{equation}
that is best interpreted as the {\it renormalized} effect of the total
mass of hydrophobic material deposited on the surface, at the
(unresolved) microscopic level.

Second, we also present results as a function of the non-homogeneity
of the hydrophobic pattern. This non-homogeneity, or {\it roughness},
can be taken as the spatial variance of the slip function:
\begin{equation}
  \Delta^2  =  \la (s(\br_w) -s_{av})^2 \ra = \frac{1}{S} \int  (s(\br_w)-\la s\ra)^2 dS.
\label{eq:sigma}
\end{equation}
In order to quantify the gain or the loss in the slip flow with
respect to the homogeneous situation, we shall focus our attention
mainly on the simplest non-trivial inhomogeneous boundary
configurations sketched in Fig. (\ref{fig:configuration}).

This corresponds to a periodic array of two strips. In the first
strip, of length $H$, the slip coefficient is chosen as
$s(\br_w)=s_{1}$.  In the second strip (of length $L-H$), we impose
$s(\br_w)=s_{0}$. We will distinguish the two cases when the strips
are oriented longitudinally or transversally to the mean flow.  In
these configurations, the total mass $s_{av}$ is given by:
$$
s_{av}=\xi s_{1} + (1-\xi)s_{0} 
$$ 
and the degree of non-homogeneity, or roughness, by:
$$
\Delta^2=\xi(1-\xi)(s_{1}-s_{0})^2. 
$$ 

By choosing (without loss of generality) $s_{1}>s_{0}$, in this
configuration the quantity $\xi=H/L$ is a natural measure of the
localization of the slip effect. This geometry allows us to compare
our results with some analytical, numerical and experimental results
for the small Knudsen regime and also to extend the study to the {\it
transition regime}.  In sub-section (\ref{sec:mean}) we shall also
present results with slightly more complex boundary conditions, namely
for the case of a bi-periodic pattern of alternating slip and no-slip
boundary conditions.

\subsection{Exact Results and Knudsen effects}
As a validation test, we first check whether our model can reproduce
some of the existing results concerning the slip properties of
hydrodynamic systems with boundaries made up of alternating strips of
zero-slip and infinite-slip lengths.

 \cite{Phi1} analyzed this situation using the
Navier-Stokes equations for the case of a cylinder with boundaries
made up of alternating longitudinal strips of perfect-slip and
no-slip.  He obtained the following exact result: \be
\label{eq:phi}
\ell_s^{long} \equiv \frac{L_s}{L_y} = \frac{2}{\pi} \log{\left(1/cos(\pi
\xi/2)\right)} \ee 

where $\xi$ is the fraction of the plate where the slip length is
infinite and where we have defined $\ell_s^{long}$ as the normalized
(to the pattern dimension) macroscopic slip length.

Notice that the r.h.s. of (\ref{eq:phi}) is independent of the radius
of the cylinder, and therefore Philip's result is directly applicable
to our geometry of Fig. (\ref{fig:configuration}), in the limit of
small Knudsen numbers.  More recently,  \cite{stone},
analyzed the same situation with the only variant of using transversal
rather than longitudinal strips.  In the limit of a cylinder with
infinite radius (plane wall boundaries), their result for the
normalized slip length can be written as:

\be 
\ell_s^{trans} \equiv
\frac{L_s}{L_x} =\frac{1}{\pi} \log(1/cos(\pi \xi/2)).
\label{eq:lauga}
\ee 

In our language, local infinite (zero) slip lengths can be obtained by
choosing $s_{1}=1$ ($s_{0}=0$).  A consistency check for our
mesoscopic H-LBE model is to reproduce the hydrodynamic limits studied
in the aforementioned papers, in the limit of small Knudsen numbers
and large channel aspect-ratio, $L_z/L_x$.

\begin{figure}
\begin{center}
\includegraphics[scale=.7]{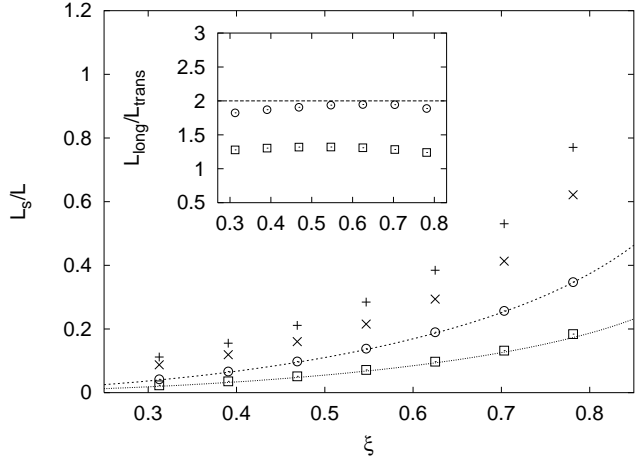}
\caption{Normalized slip length for transversal and longitudinal
  strips with $s_{1}=1$, $s_{0}=0$. We plot the normalized slip length
  as a function of the slip percentage $\xi$. The system's dimensions
  are those of Fig. (\ref{fig:configuration}). A first set of LBE
  simulation is carried out at small Knudsen, $Kn=1.10^{-3}$ for
  transversal ($\square$) and longitudinal strips ($\circ$). These
  results are compared with the analytical estimates of \cite{Phi1}
  (dashed line) and \cite{stone} (continuous line). Notice the perfect
  agreement with the analytical results in the hydrodynamic limit.
  Another set of simulations is carried out with much larger Knudsen,
  $Kn=5.10^{-2}$ to highlight the effect of rarefaction on the system
  for both traversal ($\times$) and longitudinal ($+$) strips. In the
  inset we show the ratio between the slip lengths for parallel and
  longitudinal strips for $Kn=1.10^{-3}$ ($\circ$) and $Kn=5.10^{-2}$
  ($\square$). Here we notice how by increasing the Knudsen number the
  orientation of the strip region with respect to the mean flow
  becomes less important.}
\label{fig:stonelauga}
\end{center}
\end{figure}

To this purpose, we performed a direct numerical simulation of the
H-LBE model for a channel with square cross-section, $L_x=L_y$, and
different heights, $L_z$.  For small and fixed Knudsen number, by
increasing the aspect ratio $L_{z}/L_{x}$ at fixed channel length,
$L_x$, the previous hydrodynamic limits are attained and the
normalized slip lengths $\ell^{trans}_{s}$,$\ell^{long}_{s}$ are
independent on $L_{z}$. In Fig. (\ref{fig:stonelauga}) we present the
results obtained for both longitudinal and transversal strips compared
with the analytical predictions (\ref{eq:phi}-\ref{eq:lauga}) for a
given channel aspect-ratio.

The result (see Fig. (\ref{fig:stonelauga})) shows that the analytical
hydrodynamic results are well reproduced by our mesoscopic model.
Moreover, we can go beyond the hydrodynamical limit studied by  \cite{Phi1,Phi2,stone}, and
investigate the effect of larger Knudsen numbers on these
configurations, both in the near-hydrodynamic and in the {\it
transition} regimes observed in the experiments, \cite{tabe}.  The
result (see Fig. (\ref{fig:stonelauga})) is that an increase of the
Knudsen number leads to an increase of the slip length, without
preserving the ratio between $\ell_s^{long}$ and $\ell_s^{trans}$ (see
inset of Fig. (\ref{fig:stonelauga})).  These results can be explained
by observing that upon increasing the Knudsen number, even the
'non-conductive' strips which had zero-slip length in the hydrodynamic
regime, acquire a non-zero slip due to effects of order $Kn^2$ in the
boundary conditions, \cite{ss04}.  As a result, the local slip length
(no longer equal to zero) is incremented, thereby yielding a net gain
in the overall slippage of the flow.  Let us notice that at still
relatively small Knudsen numbers, $Kn =0.05$, a fairly substantial
increase of the slip length is observed, which may reach $60-80\%$ of
the typical pattern dimension for a percentage of slipping surface
$\xi \sim 0.8$.

Another interesting question concerns the dependency of the local
velocity profile on the {\it local} slip properties with changing
Reynolds and Knudsen numbers.
\begin{figure*}
\begin{center}
\includegraphics[scale=.7]{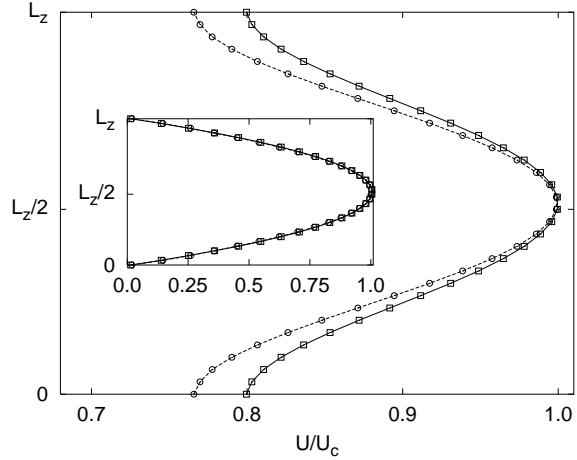}
\caption{Local velocity profile in the middle of a slip strip (the one
  with $s_{1}=1$) for transversal strips in the geometry depicted in
  Fig. (\ref{fig:configuration}).  We plot the velocity in the
  stream-wise direction as a function of the height $z$ for two
  different Reynolds numbers $Re \sim 4.5$ $(\square)$, and $Re \sim
  9.5$ $(\circ)$. The $Re$ numbers are estimated as the ratio between
  the center channel velocity of the integral profile and the sound
  speed velocity $c_{s}$. Both velocity fields are normalized with the
  center channel velocity.  The Knudsen numbers are $Kn=0.01, 0.005$
  respectively. Inset: the same but in the middle of a no-slip strip
  ($s_0=0$)}
\label{fig:onoff}
\end{center}
\end{figure*} 
We choose a transversal periodic array of strips with $H=L_{x}/2$ and
$s_{0}=0$, $s_{1}=1$ and look at the profiles in the middle of the
region with $s_{1}=1$ and in the middle of the region with $s_0=0$.
The DNS results (Fig. (\ref{fig:onoff})) clearly indicate a dependency
on the Knudsen and Reynolds numbers only in the slip region.  This is
readily understood by observing that the Reynolds number is given by
$Re=Ma/Kn$, so that, by fixing the Mach number and varying the
Reynolds number, we change also the Knudsen number, thus affecting the
local slip properties of the flow.
The most interesting result here is the inversion of concavity for the
local profile nearby the wall in the slip region: a clear indication
of the departure from the parabolic shape of the Poiseuille flow.

\begin{figure*}
\includegraphics[scale=0.55]{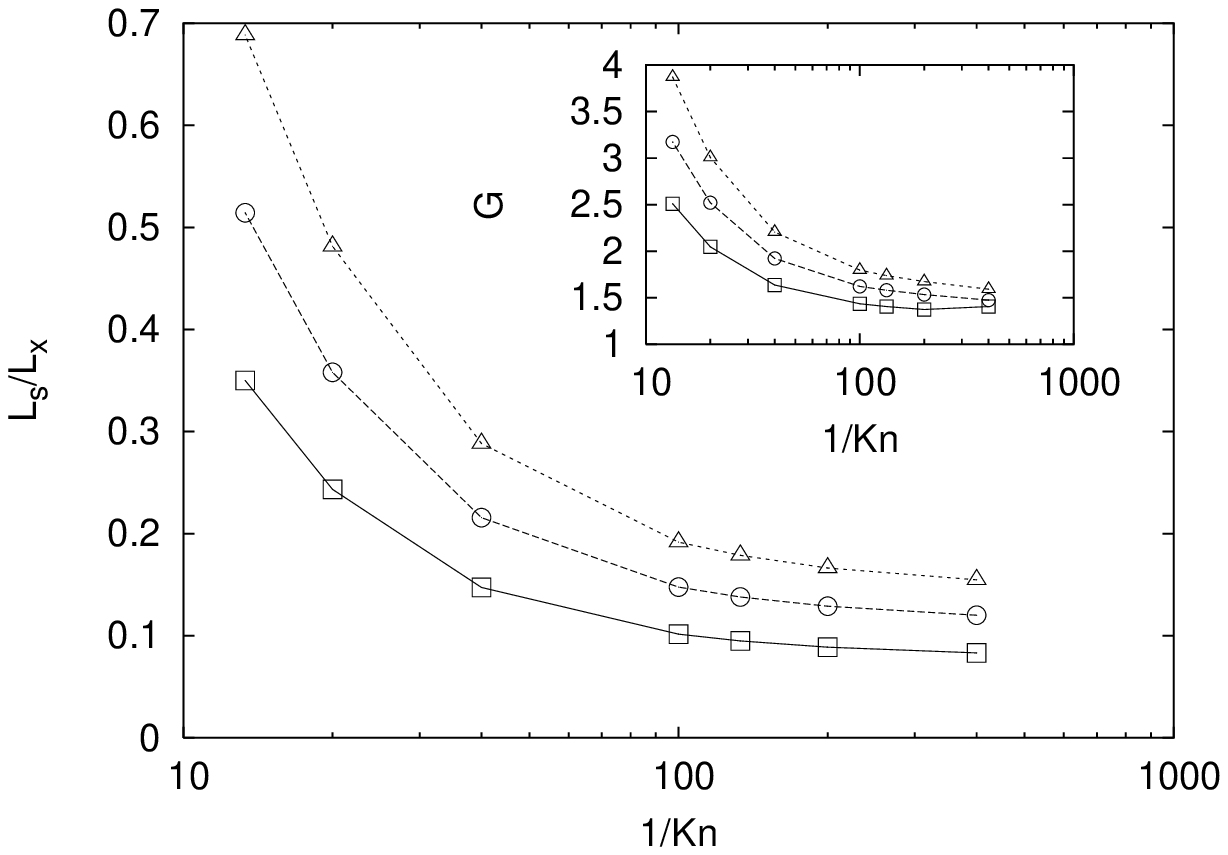}
\includegraphics[scale=0.55]{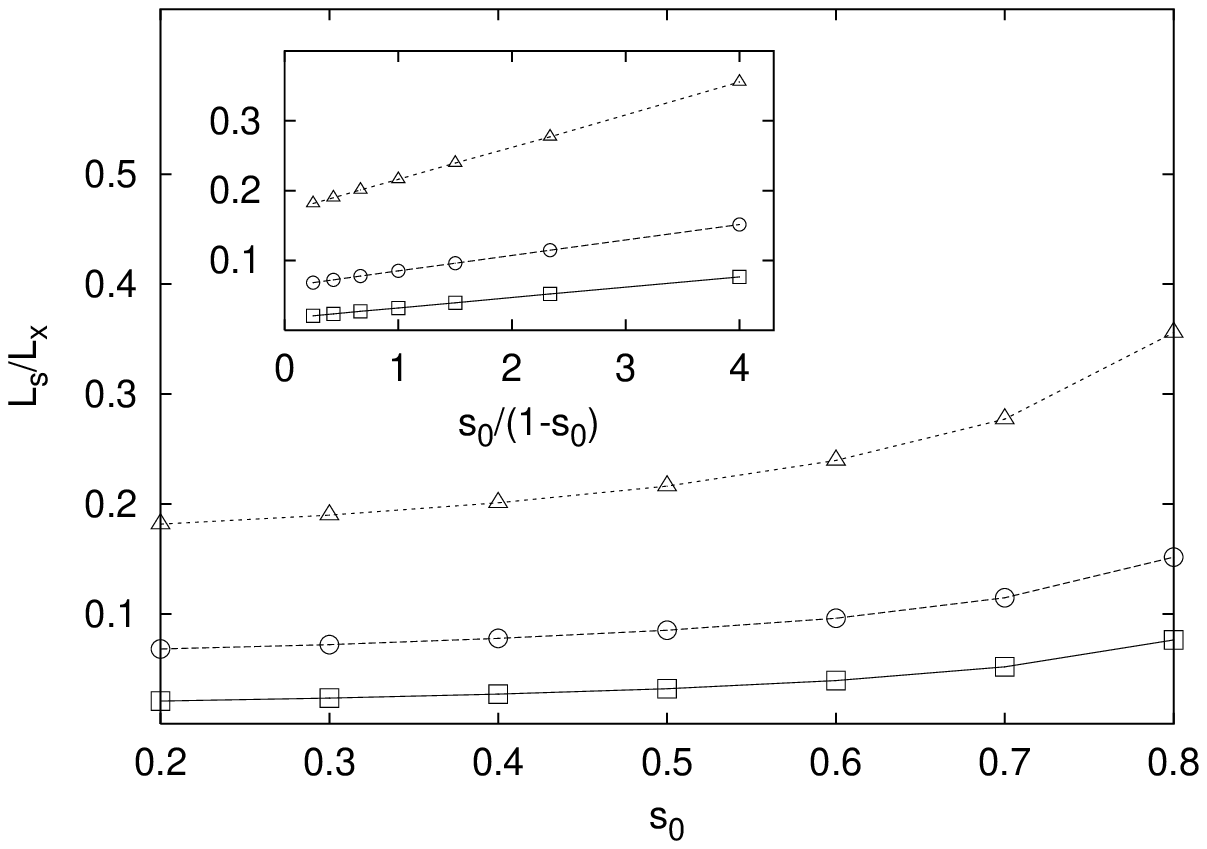}
\caption{Left panel: Normalized transversal slip length,
  $\ell^{trans}_s=\frac{L_{s}}{L_{x}}$, as a function of the average
  pressure in the system (inverse of the Knudsen number) and for
  different values of the localization parameter: $\xi=0.58$
  ($\square$), $\xi=0.65$ ($\circ$), $\xi=0.63$ ($\triangle$). The
  values of $s_{0},s_{1}$ are kept fixed to $s_0=0$ and $s_1=1$. This
  behavior is qualitatively similar to what observed in MD simulations
  of microchannels with grooves, where the degree of slippage
  localization is governed by the width of the grooves, see \cite{barrat}.
  In the inset we show the same trends but for the mass flow rate gain
  $G$ (see eq. \ref{eq:mfr}). Right panel: $\ell_s^{trans}$ as a
  function of the local degree of slippage $s_0$ for different values
  of slippage localization, $\xi=0.25,0.5,0.75$
  ($\square$,$\circ$,$\triangle$ respectively). In the inset we show
  the dependency of the slip length, $\ell^{trans}_s$, on the
  microscopic slip properties, $s_0/(1-s_0)$, for the same values of
  $\xi$.}
\label{figbarrat}
\end{figure*}

Next, we check our method against some experimental results and MD
simulations.  For example, in Fig. (\ref{figbarrat}), we show the
dependency of the transversal normalized slip length,
$\ell^{trans}_s$, as a function of the inverse of the Knudsen number,
i.e. as a function of the {\it mean channel pressure}, for different
values of the localization parameter $\xi$. This is a direct
comparison with the results in Fig. 6 of \cite{barrat} where the
evolution of the slip length as a function of the Pressure in MD
simulations of a channel with grooves of different width is
shown. Also in that case, the slip length increases by either
decreasing the pressure (increasing Knudsen) or increasing the groove
width (increasing the region with {\it infinite} slip). The two
behaviors are qualitatively similar, with a less pronounced slip
length for our case also due to the fact that we show the case of
transversal strips while in Fig. 6 of \cite{barrat} only the case of
longitudinal grooves are presented. In the right panel of
fig. (\ref{figbarrat}) we plot $\ell^{trans}_s$ at varying the level
of slippage, $s_0$, of one of the two strips (the other being kept
fixed to $s_1=1$). This is meant to investigate the sensitivity of the
macroscopic observable to the microscopic details. As one can see, the
change is never dramatic, at least for this configuration. In the
inset of the same figure one notice a linear dependency between
$\ell^{trans}_s$, and the {\it local} slip properties,
$s_0/(1-s_0)$. For {\it local} slip properties we mean the local slip
length as defined from the local boundary condition, $ \bu_{||}(\br_w)
\propto \frac{s(\br_w)}{1-s(\br_w)} |\partial_n \bu_{||}(\br_w)|$. The
same linear trend is observed in fig.  12 of \cite{barrat} using a
hydrodynamic model with suitable boundary conditions.

\begin{figure*}
\includegraphics[scale=0.4]{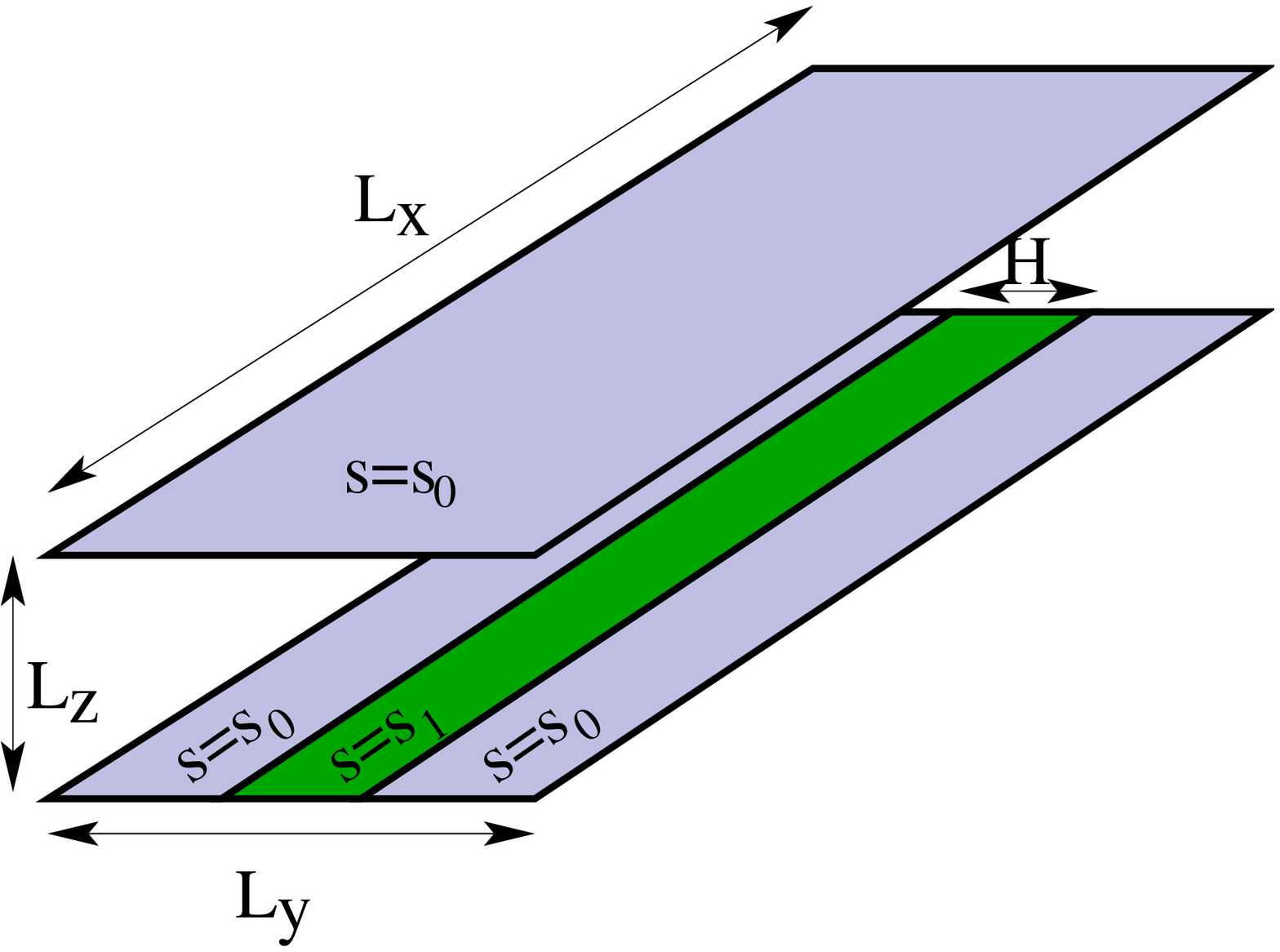}
\includegraphics[scale=0.55]{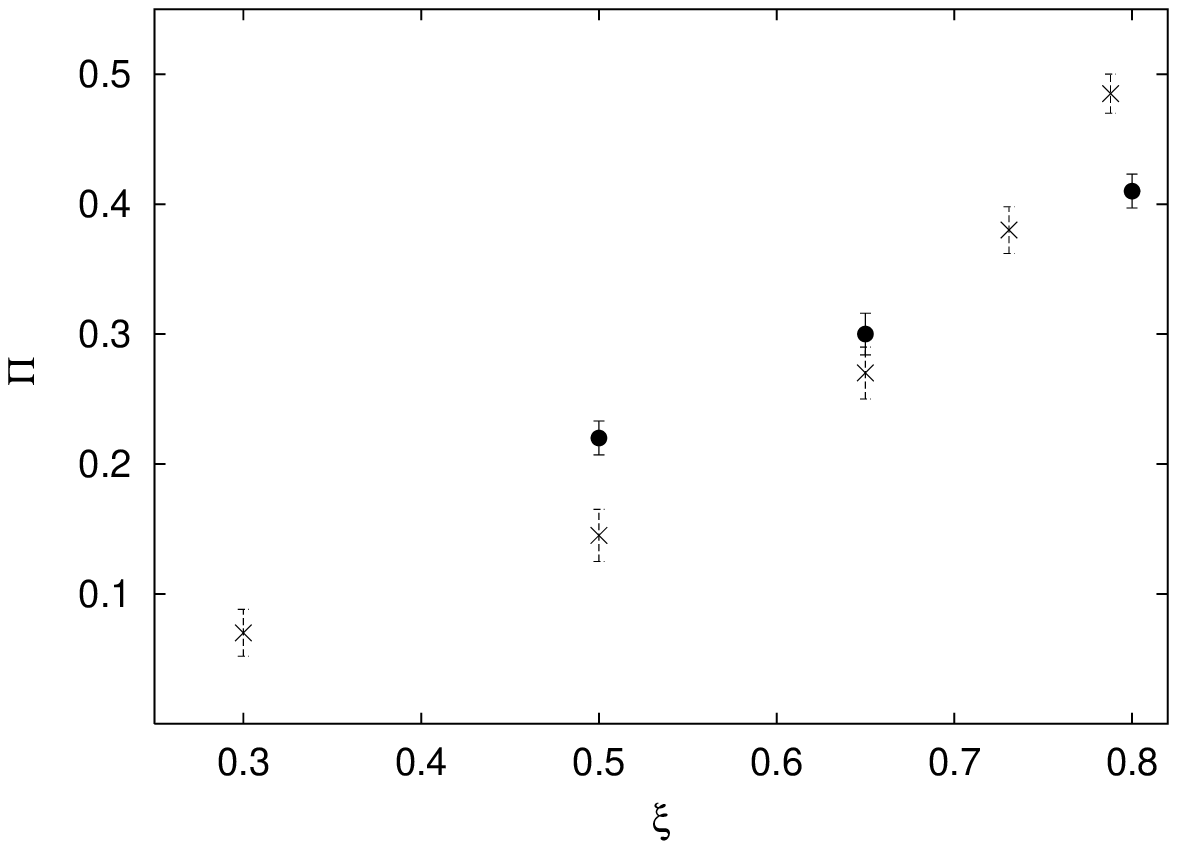}
\caption{Results for the pressure drop reduction $\Pi$ (\ref{eq:pdr}),
  as a function of the percentage of the free slip area on the
  surface.  The geometry is the same investigated experimentally in
  Fig. 15 of \cite{ou} where a micro-channel with only one surface
  engraved with longitudinal strips with super-hydrophobic material is
  studied (left panel). Here we present the results from the
  experiments ($\bullet$), superposed with our numerical simulations
  ($\times$) obtained at $Kn=5.10^{-4}$.  The numerical mesh is such
  to mimick the same channel height of $127 \mu$m used in the
  experiment. The error bars in the LBE numerics represent the maximum
  variation obtained by fixing $s_1=1$, in the shear-free area, and
  changing $s_0 \in [0.4,0.65]$ in the normal surface.  }
\label{figou}
\end{figure*}

In Fig. (\ref{figou}) we present the same kind of plot shown in the
experimental investigation   (see Fig. 15 of \cite{ou}).
Here, we plot the pressure drop reduction, $\Pi$, in the microchannel
as a function of the percentage, $\xi$, of the free slip area on the
surface (super-hydrophobic material).  We note a remarkable agreement
with the experimental results over a wide range of $\xi$, i.e. the
ratio between the regions with super-hydrophobic and normal material
on the wall.  The geometry and Knudsen number are the same of the
experiment. The only free parameters are the values of $s_0$ and $s_1$
assigned to the two different strips. Here, we have fixed $s_1=1$ in
the super-hydrophobic area, and we have varied $s_0 \in [0.4,.65]$ for
the normal material.  Notice that the LBE results exhibit the same
trend of the experiments as a function of $\xi$ and they are even in
good quantitative agreement for $\xi \sim 0.6$.  Overall, there is a
small dependency of $\Pi$ on the unknown value of $s_0$, at least in
the range considered, as already shown by the data presented in
Fig. (\ref{figbarrat}). Once the correct values of $s_1$ and $s_0$
able to reproduce the experimental results are identified, one may
easily use the present LBE method to predict and extend the outcomes
of other experiments with different geometries and/or distributions of
the same hydrophobic material on the surface.

\subsection{Effects of roughness and of  localization}
As a next step, we address of the effect of the roughness, the total
mass of the hydrophobic material, and the localization over the
surface.

To this purpose, we choose a transversal configuration where
$H=L_{x}/2$ (fixed localization, $\xi=0.5$) and
$$s_{0}=s_{av}+\Delta, \;\;\; s_{1}=s_{av}-\Delta,$$ thus yielding
$\langle s \rangle=s_{av}$ for any degree of the roughness, $\Delta$.

\begin{figure}
\begin{center}
\includegraphics[scale=.7]{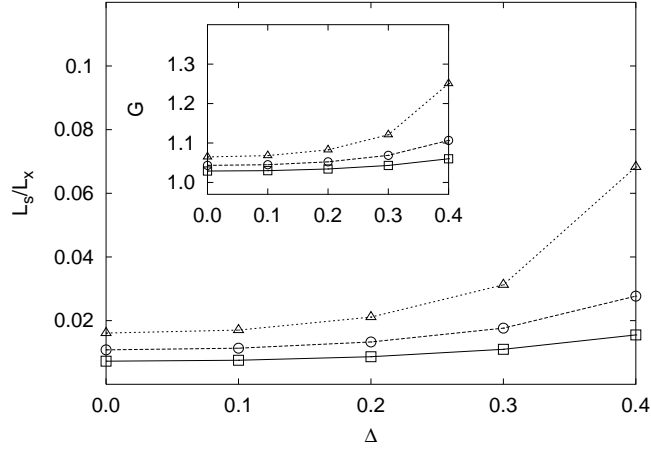}
\caption{Normalized transversal slip length, $\ell^{trans}_s =
  L_s/L_x$, for boundary conditions as in
  Fig. (\ref{fig:configuration}) with $s_{1}=s_{av}+\Delta$ in a
  fraction $\xi =H/L_{x}=1/2$ and $s_{0}=s_{av}-\Delta$ in the
  other. The Knudsen number is $Kn=0.001$.  We plot the normalized
  slip length as a function of the roughness parameter $\Delta$ and
  the following values of $s_{av}$ are considered: $s_{av}$=0.4
  ($\square$), $s_{av}=0.5$ ($\circ$), $s_{av}=0.6$ ($\triangle$).
  Inset: Same trends of the main body but for the mass flow rate gain
  $G$.}
\label{fig:roughness}
\end{center}
\end{figure}

We then look, for a given $s_{av}$, at how the slip properties of the
system respond to changes in the excursion, $\Delta$, at fixed
localization $\xi$.  The results for the normalized transversal slip
length, $\ell^{trans}_s$, and the mass flow rate gain, are presented
in (Fig. (\ref{fig:roughness})).  Both the slip length and the mass
flow rate increase by increasing $\Delta$. Notice that one can easily
reach slip lengths which are of the order of $10 \%$ of the channel
pattern dimension by increasing the roughness at fixed total mass of
slip material deposited on the surface. Similarly, the mass flow rate
gain, $G$, is increased of the order of $20-30 \%$ with respect to the
Poiseuille flow. In other words, the best throughput is obtained by
increasing the inhomogeneity of the slippage material deposited on the
surface (experimentally this means to keep the region covered by
hydrophobic molecules as segregated as possible from the region
covered with hydrophilic molecules).

\begin{figure}
\begin{center}
\includegraphics[scale=.7]{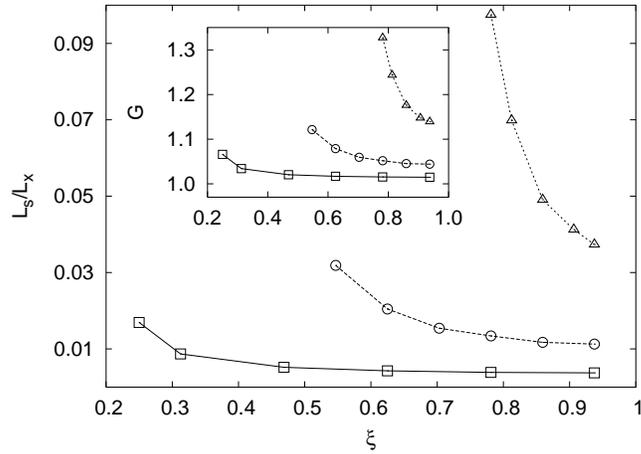}
\end{center}
\caption{Normalized transversal slip length, $\ell^{trans}_s =
  L_s/L_x$, for boundary conditions of transversal strips as in Fig.
  (\ref{fig:configuration}) with $s_{1}=s_{av}/\xi$ in a fraction
  $\xi$ and $s_{0}=0$ in the other. The Knudsen number is $Kn=0.001$.
  We plot the normalized slip length as a function of the localization
  parameter $\xi$ for the following values of $s_{av}$: $s_{av}=0.25$
  ($\square$), $s_{av}=0.5$ ($\circ$), $s_{av}=0.75$ ($\triangle$).
  Inset: Same trends of the main body but for the mass flow rate gain
  $G$.}
\label{figlocalization}
\end{figure}

Another possible way to compute the slippage effects is to analyze the
slippage at fixed total mass of hydrophobic material, varying both the
localization and the roughness. To this purpose, we choose again a
transversal configuration where we fix the total mass $s_{av}$ and we
choose $s_{1}=s_{av}/\xi$, $s_{0}=0$, $\xi=H/L_{x}$ being the degree
of localization associated with a given roughness. The result
(Fig. (\ref{figlocalization})) is that the slippage is greater as the
the degree of localization, and --consequently-- of roughness, is
increased.

\subsection{Mean field approach and beyond}
\label{sec:mean}
We observe that a mean field approach is able to reproduce the {\it
qualitative} trends observed so far.

In the boundary condition (\ref{BBB}) there is a coupling between the
local velocity field and the stresses at the wall. Obviously, the
averaged slip length depends both on $s(\br_w)$ and on the stresses.

In order to highlight the effect of the slip function on the mean
quantities, as a first approximation, we can leave the wall stress
fixed at its Poiseuille value, and work only on the properties of
$s(\br_w)$, namely: \be \langle u_{slip} \rangle \sim \langle
\frac{s}{1-s} \rangle. \ee For the configuration analyzed so far,
without loss of generality we define:
$$\delta=\frac{s_{1}-s_{0}}{2} \hspace{.2in}
s_{+}=\frac{s_{0}+s_{1}}{2}$$
and write the averaged slip properties
$\langle \frac{s}{1-s} \rangle$ as a function of $\delta$ and $s_{+}$:
\be \langle \frac{s}{1-s} \rangle =
p_{0}\frac{s_{0}}{1-s_{0}}+p_{1}\frac{s_{1}}{1-s_{1}} =
p_{0}\frac{s_{+}-\delta}{1-s_{+}+\delta}+p_{1}\frac{s_{+}+\delta}{1-s_{+}-\delta}
\ee 

where $p_{1}=\frac{H}{L_{x}}$, $p_{0}=1-p_{1}$ are the percentages of
the surface associated with slip and no-slip areas respectively.
Making use of Taylor expansion up to second order in $\delta$ we get:

\be \langle \frac{s}{1-s} \rangle \approx
\frac{s_{+}}{1-s_{+}}+\frac{\delta}{(1-s_{+})^{2}}(p_{1}-p_{0})+\frac{\delta^{2}}{(1-s_{+})^{3}}.
\ee Since $p_{1}=\frac{H}{L_{x}}=\xi$ and $p_{0}+p_{1}=1$, we finally
obtain: \be \langle \frac{s}{1-s} \rangle \approx
\frac{s_{+}}{1-s_{+}}+ \frac{\delta (2 \xi-1)(1-s_{+}) +
\delta^{2}}{(1-s_{+})^{3}}.  \ee

First, in our case of a fixed localization, by setting $\xi=1/2$ we
have $s_{+}=s_{av}$, $\delta=\Delta$ and we obtain: \be \langle
\frac{s}{1-s} \rangle \approx \frac{s_{av}}{1-s_{av}}+
\frac{\Delta^{2}}{(1-s_{av})^{3}} \ee that results in a greater
slippage when the roughness $\Delta$ is increased.

Second, if we choose $s_{1}=\frac{s_{av}}{\xi}$ and $s_{0}=0$, as for
the case with fixed total mass, we obtain $\delta=\frac{s_{av}}{2\xi}$
and $s_{+}=\frac{s_{av}}{2 \xi}$ . This results in \be \langle
\frac{s}{1-s} \rangle \approx \frac{s_{av}}{(2 \xi-s_{av})}+
\frac{s_{av}(4\xi^2 -2 \xi)(2 \xi - s_{av})+2 \xi s^{2}_{av} }{(2 \xi
- s_{av})^{3}} \ee that, as a function of the localization $\xi$
yields a qualitative agreement with our analysis, supporting the idea
that the effect of slippage is greater when slip properties are
localized.

\begin{figure*}
\includegraphics[scale=0.4]{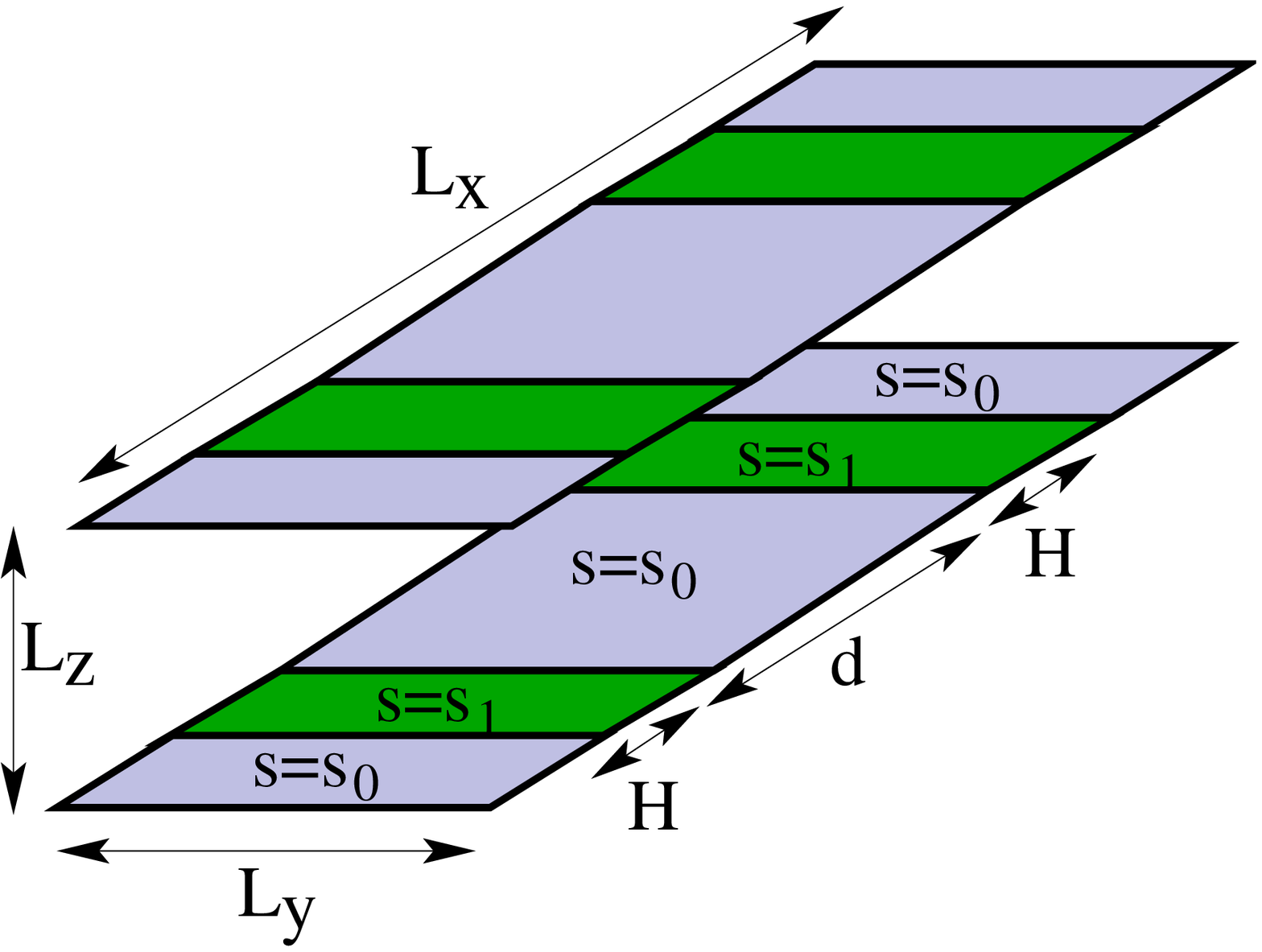}
\includegraphics[scale=.55]{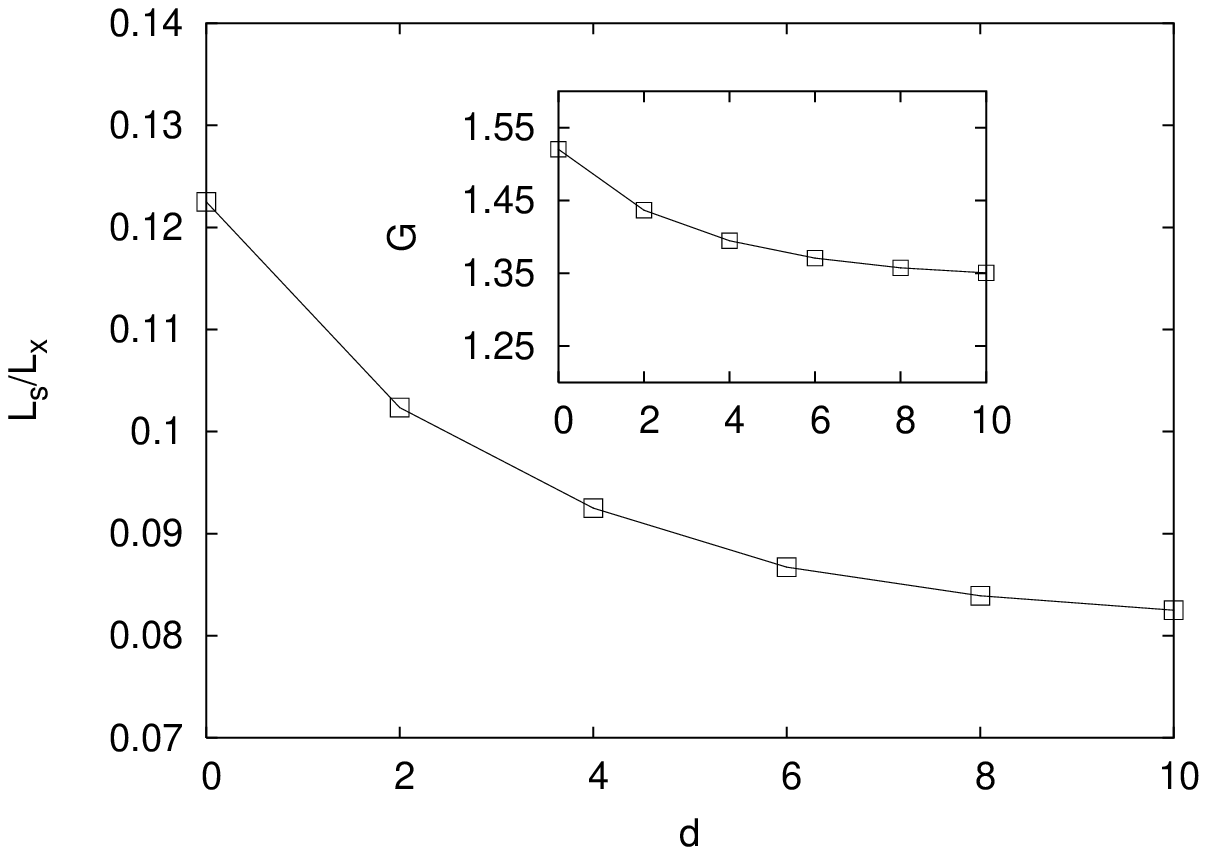}
\caption{Left: The configuration with a transversal bi-strip structure
  at the walls.  The total boundary lengths are $L_{x},L_{y}$ (stream,
  span).  The slip coefficient is chosen as $s_1=1$ in two strip of
  length $H$ and $s_0=0$ in the others. The distance between the two
  strips is $d$.  Periodic boundary conditions are always assumed in
  the span-wise and stream-wise direction. Right: results for the slip
  length and the mass flow rate gain (inset) as a function of the
  distance $d$ (in lattice units) between the two free-shear ($s=s_1$)
  strips of width $H=20$ (in lattice units). All the other parameters,
  $L_x=L_y=64, L_z=84, Kn=1.10^{-3}$ are kept fixed. }
\label{fig:biconfiguration}
\end{figure*}

It should be appreciated that the mean field approach discussed above
is not exhaustive.  In fact, we can design an experiment with the
boundary configuration sketched in Fig. (\ref{fig:biconfiguration}),
and investigate the total slippage as a function of the distance, $d$,
between the strips.  For this geometry, the mean field approach
presented before would yield the same results irrespectively of $d$.

On the other hand, we expect non-linear effects to be present when the
strips get close enough, due to the correlation between $s(\br_w)$ and
the stress at the boundary, $\partial_n u(\br_w)$.
Indeed, as one can see in Fig. (\ref{fig:biconfiguration}), the
slippage is increased when the two strips get closer to each other.

This effect, even if only of the order of $10 \%$ in the mass flow
rate with respect to the configuration for $d \gg 1$, cannot be
captured by the previous {\it mean field} argument.  Let us notice
that a similar sensitivity to the geometrical pattern of the slip and
no-slip areas has been recently reported in the experimental
investigation of \cite{ou}, where it is found that for the same
microchannel geometries and shear-free area ratios, microridges
aligned in the flow direction consistently outperform regular arrays
of microposts. Similar considerations have been also presented by
\cite{vinorev}.

\section{Conclusions}
\label{sec:conc}
We have presented a mesoscopic model of the fluid-wall interactions
which proves capable of reproducing some properties of flows in
microchannels. We have defined a suitable implementation of the
boundary conditions in a lattice version of the Boltzmann equation
describing a single-phase fluid in a microchannel with heterogeneous
slippage properties on the surface.  In particular, we have shown that
it is sufficient to introduce a slip function, $0\le s(\br_w) \le 1$,
defining the local degree of slip of mesoscopic {\it molecules} at the
surface, to reproduce qualitatively and, in some cases, even
quantitatively, the trends observed either in MD simulations or is
some experiments. The function $s(\br_w)$ plays the role of a {\it
renormalizing} factor, which incorporates microscopic effects within
the mesoscopic description.

We have analyzed slip properties in terms of slip length, $L_s$, slip
velocity, $V_s$, pressure drop gain, $\Pi$ and mass flow rate
$\Phi_s$, as a function of the degree of slippage, and its spatial
localization.  The latter parameter mimicking the degree of roughness
of the ultrahydrophobic material in real experiments.

With a proper choice of the slip function $s(\br_{w})$ in longitudinal
and transversal configurations, we have reproduced previous analytical
results concerning pressure-driven hydrodynamic flows with boundaries
made up of alternating strips of zero-slip and infinite-slip
(free-shear) lengths,  \cite{Phi1,Phi2,stone}.  We have also discussed
the increment of the slip length in the {\it transition regime},
i.e. where the Maxwell-like slip boundary conditions (\ref{eq:bcslip})
are supposed to be replaced by second-order ones (\ref{eq:bcslip2}).

The {\it local} velocity profile has also been studied with changing
Reynolds and Knudsen numbers and the local slip properties on the
surface.

The method introduced is able to describe slip lengths of the order of
the total height of the channel (of the order of tens of $\mu$m), or
fractions thereof. This is accompanied by an important increase in the
mass flow rate, or equivalently, in the pressure drop gain.  Whenever
possible, we have compared the results based on the Heterogeneous LBE
with MD simulations and with some recent experiments.

In particular, we have shown that the H-LBE approach is able to
reproduce the increase of the slip length as a function of the inverse
of the mean pressure in the channel, as observed in recent MD
simulations by \cite{barrat}. Concerning the same MD simulations, we have found a similar linear
dependency of the macroscopic slip lengths, $L_s$ as a function of the
microscopic slip properties at the surface. As to the experiments, we have shown that the H-LBE
approach is able to achieve quantitative agreement with the
experimental study presented in \cite{ou}, concerning the slip
properties as a function of the relative importance of regions with
high-slip and low-slip at the surface.  The natural application of our
numerical tool consists in tuning the free parameters $s_0$ and $s_1$
in order to reproduce experimental results in controlled geometries.

Then, one may use the LBE scheme with the given $s_0$ and $s_1$
values, to explore flows in different geometries and/or with different
patterns of the same slip and no-slip materials.

The method is a natural candidate to study flow properties in more
complex geometries, of direct interest for applications.  Transport
and mixing of active or passive quantities (macromolecules, polymers
etc...) can also be addressed.

By definition, the present H-LBE description is limited to a {\it
phenomenological } interpretation of the slip function.  A natural
development of this approach, is to implement a multi-phase Boltzmann
description, able to attack the wall-fluid interactions and
fluid-fluid interactions at a more microscopic level.

This route should open the possibility to discuss the formation of a
gas phase close to the wall, induced by the microscopic details of the
fluid-wall physics.  Results along this direction, will make the
object of a forthcoming publication, \cite{bbsst05}.

\section{Appendix A}

\begin{figure*}

{\hfill\includegraphics[scale=0.3]{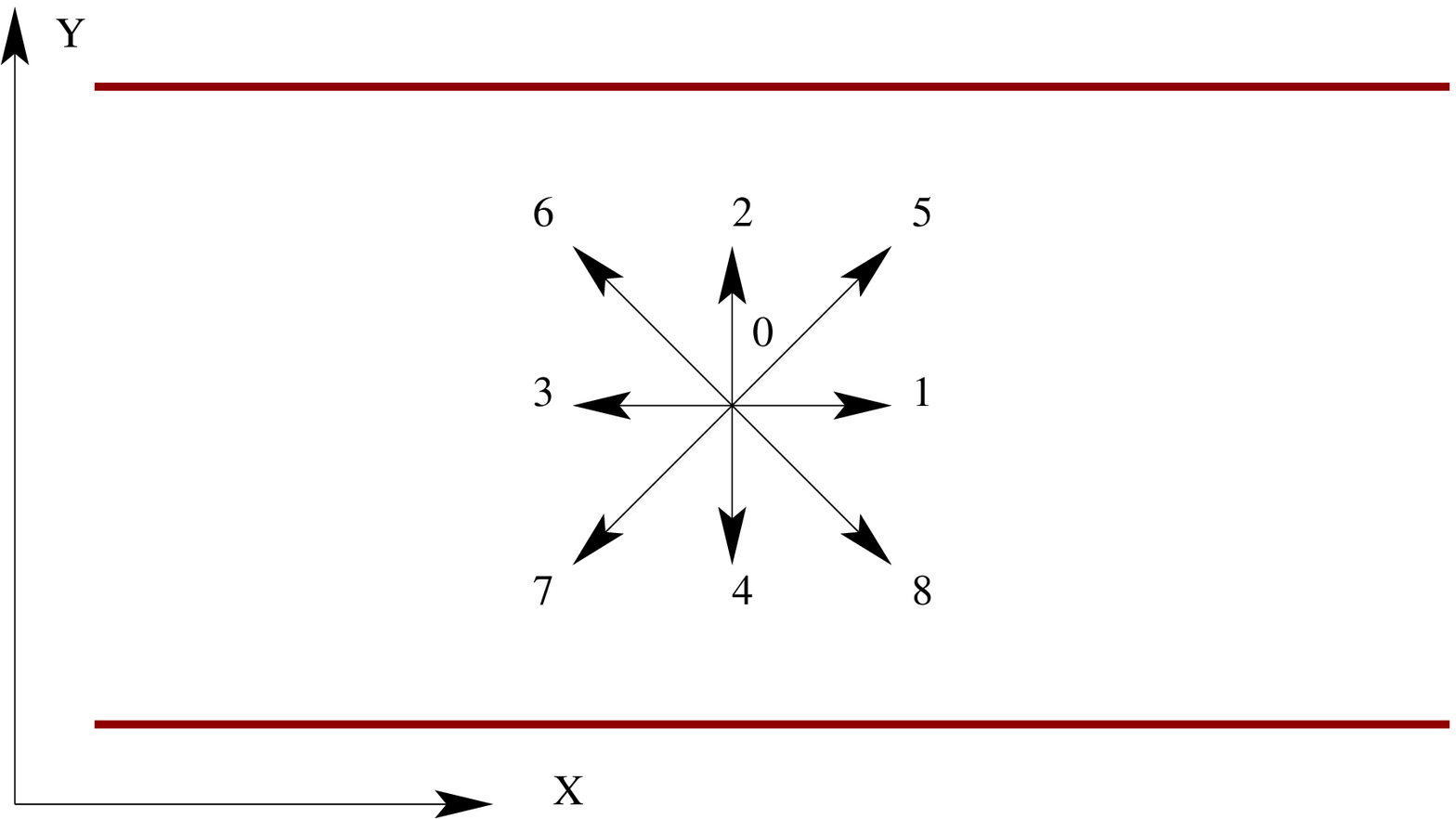} \hfill
\includegraphics[scale=0.5]{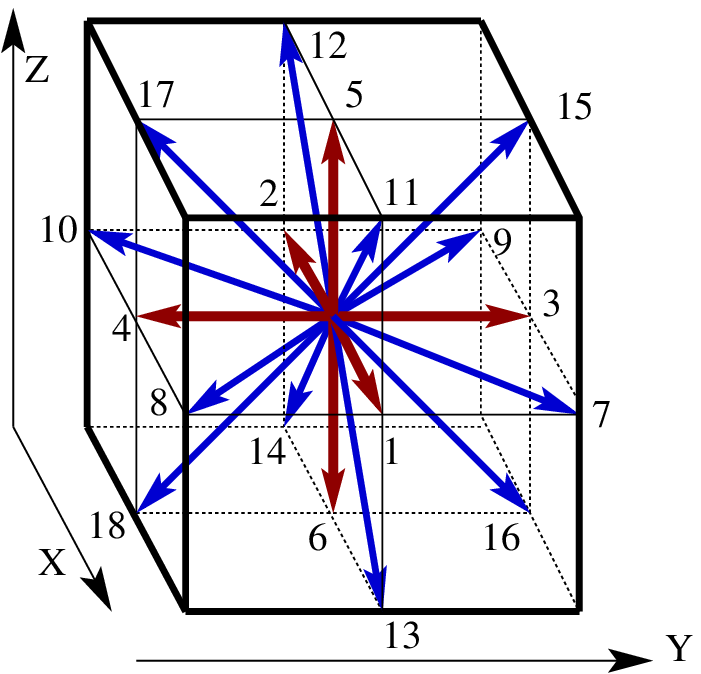} \hfill}
\caption{2d and 3d lattice discretization of the velocities in the LBE
  schemes used in this work. The velocities entering in the north wall
  in the 2d scheme (left) are, $f_5,f_2,f_6$. The outgoing velocities
  are: $f_7,f_4,f_8$. At the south wall the roles are exchanged.}
\label{fig00}
\end{figure*} 

The Lattice Boltzmann Equation (LBE) for a Pressure-driven channel
flow is a streaming and collide equation involving the particle
distribution function $f_{l}(\vec{x},t)$ of finding a particle with
velocity $\vec{c}_{l}$ (discrete velocity phase space) in $\vec{x}$ at
time $t$. The equation is written in the following form:
\be\label{fullLBE}
f_{l}(\vec{x}+\vec{c}_{l}\dt,t+\dt)-f_{l}(\vec{x},t)=
-\frac{1}{\tau}\left(f_{l}(\vec{x},t)-f^{(eq)}_{l}(\rho,\vec{u})\right)+\frac{\dx}{c^{2}}
F g_{l} \ee with $\tau$ the relaxation time and $g_{l}$ the forcing
term projection with the property \be \sum_{l} g_{l}=0 \hspace{.2in}
\sum_{l} g_{l} \vec{c}_{l}= 1.  \ee For the case of two dimensional
grid ($2DQ9$) depicted in Fig. (\ref{fig00}), for example, the
$g_{l}$'s can be taken with the following properties: \be g_{1}=-g_{3}
\hspace{.2in} g_{5}=g_{8}=-g_{6}=-g_{7} \ee leaving only one unknown
parameter, say $g_{5}$. Discrete space and time increments are
$\dx,\dt$, with $c=\frac{\dx}{\dt}$ the intrinsic lattice velocity.
The equilibrium distribution $f^{(eq)}_{l}(\rho,\vec{u})$ is given by:
\be\label{fullequi} f^{(eq)}_{l}(\rho,\vec{u})=w_{l} \rho
\left[1+\frac{(\vec{c_{l}} \cdot
    \vec{u})}{c_s^{2}}+\frac{1}{2}\frac{(\vec{c_{l}} \cdot
    \vec{u})^{2}}{c_s^{4}}-\frac{1}{2}\frac{u^{2}}{c_s^{2}}\right] \ee
being $c^2_{s}=\frac{1}{3} c^2$ the sound speed velocity.  Concerning
the $2DQ9$ model here used for the technical details, the velocity
phase space is identified by the following discrete set of velocities:
\be \vec{c}_{\alpha}=\left \{ \begin{array}{c c c}
    \vec{c}_{0} & = & (0,0)c \\
    \vec{c}_{1},\vec{c}_{2},\vec{c}_{3},\vec{c}_{4} & = &
    (1,0)c,(0,1)c,(-1,0)c,(0,-1)c \\
    \vec{c}_{5},\vec{c}_{6},\vec{c}_{7},\vec{c}_{8} & = &
    (1,1)c,(-1,1)c,(-1,-1)c,(1,-1)c \end{array} \right.  \ee and the
equilibrium weights are $w_{0}=4/9$, $w_{l}=1/9$ for $l=1,...,4$,
$w_{l}=1/36$ for $l=5,...,8$.\\ As far as the 3d model we use in the
numerical analysis ($3DQ19$), it is a $19$ velocity model whose
velocity phase space is identified by: \be \vec{c}_{\alpha}=\left \{
  \begin{array}{c c c}
    \vec{c}_{0} & = & (0,0,0)c \\
    \vec{c}_{1,2},\vec{c}_{3,4},\vec{c}_{5,6} & = & (\pm
    1,0,0)c,(0,\pm 1,0)c,(0,0,\pm 1)c \\
    \vec{c}_{7,...,10},\vec{c}_{11,...,14},\vec{c}_{15,...,18} & = &
    (\pm 1,\pm 1,0)c,(\pm 1,0,\pm 1)c,(0,\pm 1,\pm 1)c \end{array}
    \right.  \ee and equilibrium weights $w_{0}=1/3$, $w_{l}=1/18$ for
    $l=1,...6$, $w_{l}=1/36$, $l=8,...,19$.\\ Our hydrodynamic
    variables such as density $\rho$ and momentum $\rho \bu$ are
    moments of the distribution function $f_{l}=f_{l}(\vec{x},t)$: \be
    \rho = \sum_{l} f_{l} \hspace{.2in} \rho \bu = \sum_{l} \vec{c}
    f_{l} \ee and in order to derive Hydrodynamic equations from
    (\ref{fullLBE}), we must consider the following expansions:
    \be\label{e1}
    f_{l}(\vec{x}+\vec{c}_{l}\dt,t+\dt)=\sum_{n=0}^{\infty}
    \frac{\epsilon^n}{n !} D^{n}_{t} f_l(\vec{x},t) \ee \be\label{e2}
    f_{l}=\sum_{n=0}^{\infty} \epsilon^{n} f^{(n)}_{l} \ee
    \be\label{e3} \partial_{t}=\sum_{n=0}^{\infty} \epsilon^{n}
    \partial_{t_{n}} \ee where $\epsilon=\delta_{t}$ and
    $D_{t}=(\partial_{t}+\vec{c}_{l} \cdot \nabla)$.% Self consistency
    order by order in $\epsilon$ and (\ref{fullLBE}) imply:
% \be   \begin{array} {l l} 
% O(\epsilon^{0}): & f^{(0)}_{l}=f^{(eq)}_{l}         \\
% O(\epsilon^{1}): &  D_{t_{0}}f^{(0)}_{l}=-\frac{1}{\tau} f^{(1)}_{l}        \\
% O(\epsilon^{2}): &  \partial_{t_{1}} f^{(0)}_{l}+ \left( \frac{2\tau-1}{2\tau} \right) D_{t_{0}} f^{(1)}_{l} =-\frac{1}{\tau} f^{(2)}_{l}.    \end{array}    
%\ee 
%The distribution function is then constrained by the following relations:
%\be
%      \sum_{l} f^{(0)}_{l} \left [  \begin{array}{l} 1 \\ 
%                                \vec{c}_{l} \end{array} \right ] = \sum_{l} f^{(eq)}_{l} \left [  \begin{array}{l} 1 \\ 
%                                \vec{c}_{l} \end{array} \right ] =                   \left [  \begin{array}{l} \rho \\ 
%                                                                                     \rho \bu \end{array} \right ] . 
%\ee
%\be
% \sum_{l} f^{(n)}_{l} \left [  \begin{array}{l} 1 \\ 
%                                \vec{c}_{l} \end{array} \right ] =  0 \hspace{.2in} n>0. 
%\ee
 We can use the expansions (\ref{e1}),(\ref{e2}),(\ref{e3}) in
(\ref{fullLBE}) and by equating order-by-order in $\epsilon$
self-consistent constraints on $f^{(n)}_{l}$ are obtained.

Up to the first order in $\epsilon$ with
$\partial_{t}=\partial_{t_{0}}+ \epsilon \partial_{t_{1}}$ we obtain
the following equations: \be \left \{ \begin{array} {l} \partial_{t}
\rho + \nabla (\rho \bu) = 0 \\ \partial_{t} \bu + (\bu \cdot \nabla
)\bu = - \frac{1}{\rho}{\bf \nabla }P + \frac{1}{\rho} \nabla \cdot
(\nu \rho \nabla \bu) \end{array} \right.  \ee with
$\nu=\frac{(\tau-\frac{1}{2})}{3}\frac{\delta^{2}_{x}}{\delta_{t}} $
and where ${\bf \nabla} P$ contains both the imposed mean pressure
drop, ${\bf F}$, and the fluid pressure fluctuations.

\section{Appendix B}
Let us now go back to eq. (\ref{fullLBE}), and derive explicitly the
non-homogeneous boundary conditions used in the text (\ref{BBB}), in
the limit $\delta_x = \delta_t \rightarrow 0$, with
$c=\frac{\delta_{x}}{\delta_{t}} \rightarrow 1$. We specialize to the
{\it steady-state} boundary condition at the north-wall ($z=L_z$) for
the $2d$ lattice ($2DQ9$):
\begin{eqnarray}
f_{7}(\br_{w})&=&(1-s(\br_{w})) f_{5}(\br_{w}) + s(\br_{w})
f_{6}(\br_{w}) \nonumber \\
f_{4}(\br_{w})&=&f_{2}(\br_{w}) \nonumber \\
f_{8}(\br_{w})&=&(1-s(\br_{w})) f_{6}(\br_{w}) + s(\br_{w})
f_{5}(\br_{w}). \nonumber
\end{eqnarray}
Assuming a constant density profile $\rho=1$ in the fluid, by
definition we have for $\br=\br_{w}$ : \be
u_{||}(\br_{w})=f_{1}(\br_{w})-f_{3}(\br_{w})+f_{5}(\br_{w})-f_{6}(\br_{w})+f_{8}(\br_{w})-f_{7}(\br_{w}).
\ee 

In the limit of small Mach numbers, disregarding all ${\cal O}(u^{2})$
terms in the equilibrium distribution and using the steady state,
$\partial_t f=0$, expansion: 

\be
\label{bf} 
f_{l}(\br)=\sum_{n=0}^{\infty} (-1)^{n} (\tau (\bc_{l}
\nabla))^{n} [ f^{(eq)}_{l}(\rho,\bu) + \tau F_l ]
\ee 
we finally obtain the estimate for the slip velocity $u_{||}$:  
\be   \begin{array} {l l} 
u_{||}(\br_{w})=2 F \tau g_{1}+\frac{2}{3} u_{||}(\br_{w}) + \frac{2}{3} c^{2}\tau^{2}
\partial_{x} u_{||}(\br_{w})+ \\
+ 2 s [ 2 F \tau g_{5}+
  \frac{u_{||}(\br_{w})}{6}- \frac{c \tau}{6} \partial_{x} u_{\bot}(\br_{w}) - \frac{c \tau}{6} \partial_{y} u_{||}(\br_{w})  +\frac{c^{2}
\tau^{2}}{6} (\partial^{2}_{x}+\partial^{2}_{y}) u_{||}(\br_{w})
+\frac{ c^{2} \tau^{2}}{3} \partial_{x} \partial_{y}  u_{||}(\br_{w}) ]+ {\cal O}(\tau^{3}). \end{array}
\label{slip}
\ee

By noticing that the external forcing, $F$, is of the order of
magnitude of the second-order stress, $|\partial^{2}_{y}
u_{||}(\br_{w})| \nu $, and $\nu = c_s^2 \tau$, the first order in
$Kn$ of (\ref{slip}) reads:

\be
\bu_{||}(\br_w)= Kn \left( \frac{c}{c_s} \right )
\frac{s(\br_w)}{1-s(\br_w)}  | \partial_n \bu_{||}(\br_w)|
\ee

where we have used $L_{z} \partial_{y}=\partial_{n}$ ,
$\tau=\frac{L_{z} Kn}{c_{s}}$, $ \partial_n \bu_{||}(\br_w)=
-|\partial_n \bu_{||}(\br_w)|$, which is the expression used in the
text.

The second-order calculation in $Kn$ is particularly simple if we
specialize to an homogeneous case ($\partial_{x} (\bullet)=0$).  After
some calculations, we obtain: \be \bu_{||} = Kn \left( \frac{c}{c_s}
\right ) \frac{s}{1-s} | \partial_n \bu_{||} | + Kn^{2} \left (
\frac{c}{c_{s}} \right )^{2} (1-4 g_{5}) | \partial^{2}_n \bu_{||}|
\ee which is the second order, in $Kn$, boundary conditions, with
unknown parameters $s$ and $g_5$ ($0 \le g_{5} \le 1/4$ ) used by
\cite{ss04}.

%\bibliography{literatur}

\end{document}